\documentclass[prb,longbibliography,twocolumn,superscriptaddress]{revtex4-2}
\usepackage{times}
\usepackage{graphicx}
\usepackage{amsmath}
\usepackage{amssymb}
\usepackage{euscript}
\usepackage{verbatim}
\usepackage{wrapfig}
\usepackage{setspace}
\usepackage{xcolor}
\usepackage{amsfonts}
\usepackage{wasysym}
\usepackage{subfigure}
\usepackage{amssymb}
\usepackage{bbm}
\usepackage{wasysym}
\usepackage[hypertexnames=false]{hyperref}

\newcommand{\spin}[1]{\mathbf{S}_{#1}}

\newcommand{\rthree}[0]{${\sqrt{3} \times \! \sqrt{3}}$ }
\newcommand{\qz}[0]{${q\!=\!0}$ }
\newcommand{\ztwo}{$\mathbb{Z}_2$ }

\begin{document}
\title{Order by disorder in classical kagom\'e antiferromagnets with chiral interactions}
\author{Jackson Pitts}
\affiliation{Department of Physics and Astronomy, University of California, Riverside,  CA 92521, USA}
\author{Finn Lasse Buessen}
\affiliation{Department of Physics, University of Toronto, Toronto, Ontario M5S 1A7, Canada}
\affiliation{Institute for Theoretical Physics, University of Cologne, Z\"{u}lpicher Stra{\ss}e 77, 50937 K\"{o}ln, Germany}
\author{Roderich Moessner}
\affiliation{Max-Planck-Institut f\"{u}r Physik komplexer Systeme, N\"{o}thnitzer Stra{\ss}e 38, 01187 Dresden, Germany}
\author{Simon Trebst}
\affiliation{Institute for Theoretical Physics, University of Cologne, Z\"{u}lpicher Stra{\ss}e 77, 50937 K\"{o}ln, Germany}
\author{Kirill Shtengel}
\affiliation{Department of Physics and Astronomy, University of California, Riverside, CA 92521, USA}
\affiliation{Max-Planck-Institut f\"{u}r Physik komplexer Systeme, N\"{o}thnitzer Stra{\ss}e 38, 01187 Dresden, Germany}

\begin{abstract}

The Heisenberg antiferromagnet on the kagom\'e lattice is an archetypal instance of how large ground state degeneracies arise, and how they may get resolved by thermal and quantum fluctuations.
Augmenting the Heisenberg model by chiral spin interactions has proved to be of particular interest in the discovery of
certain chiral quantum spin liquids.
Here we consider the classical variant of this chiral kagom\'e model and find that
it  exhibits, similar to the classical Heisenberg antiferromagnet, a remarkably large and structured ground-state manifold, which combines
continuous and discrete degrees of freedom. This allows for a rich set of order-by-disorder phenomena.
Degeneracy lifting occurs in a highly selective way, choosing already at the harmonic level  specific triaxial states which however retain an emergent \ztwo degree of freedom (absent in the conventional Heisenberg model).
We also study the competition of entropic and energetic ground state selection as the model interpolates between the purely chiral and Heisenberg cases. For this mixed model, we find a ``proximate ordered-by-disorder'' finite-temperature regime where fluctuations overcome the energetic ground state preference of the perturbation.  Finally, a semiclassical route to a spin liquid is provided by quantum order by disorder in the purely chiral models, where the aforementioned \ztwo degrees of freedom are elevated to the role of an emergent gauge field.

\end{abstract}

\maketitle


\section{Introduction}
In many-body physics, the formation and splitting of accidental degeneracies is a recurring motif in the emergence of unconventional states of matter. Examples include the fractional quantum Hall states arising from interaction effects in partially filled (and hugely degenerate) Landau levels \cite{Laughlin1983} or the plethora of correlated states observed in moire materials \cite{Balents2020,Andrei2020,Kennes2021} where, e.g., the twisting to a magic angle leads, in the non-interacting limit, to a flat, degenerate band \cite{Bistritzer2011}. The phenomenology of residual degeneracies, which are not protected by any symmetries, has also been broadly studied in the field of frustrated magnetism. Here it is the competition of interactions, often induced by the geometry of the underlying lattice, that hinders the formation of conventional magnetic order and keeps the magnetic moments fluctuating down to the lowest temperatures \cite{Wannier1950}, orders of magnitude below the Curie-Weiss scale set by the bare magnetic interactions \cite{Moessner2001,Moessner2006}.
This opens the stage for residual effects to instigate the macroscopic state of matter by lifting the degeneracy of fluctuating states.
One of the most intriguing scenarios here is the phenomenon of order by disorder \cite{Villain1980} where the fluctuations themselves
induce magnetic order by effectively lowering the free energy for a selection of states. Thermal order-by-disorder transitions to low-temperature magnetic order have been discussed for Heisenberg antiferromagnets for various lattice geometries incompatible with the formation of a simple N\'eel state such as the face-centered cubic (fcc) \cite{Henley1987,Henley1989} or pyrochlore \cite{Villain1979,Moessner1998a}.   The spinel materials whose magnetism is described by competing nearest and next-nearest neighbor couplings on the diamond lattice \cite{Bergman2007} have been particularly scrutinized \cite{Bernier2008,Chen2017,Buessen2018} in light of
two-stage magnetic ordering in MnSc$_2$S$_4$ \cite{Fritsch2004} where the \emph{upper} transition was argued to arise from thermal order-by-disorder physics -- experimental verification of which came in recent neutron scattering experiments revealing the formation of a degenerate spiral spin liquid manifold in the precursor of this transition \cite{Gao2017}.
In going one step further, the breaking of classical symmetries by {\em quantum} fluctuations is known as quantum order by disorder. This phenomenon has been extensively studied, with the pyrochlore antiferromagnet again being a principal example~\cite{Hizi2005,Hizi2006}.
An important case in point for the experimental observation of quantum order by disorder are the  inelastic neutron scattering measurements of the pyrochlore compound Er$_2$Ti$_2$O$_7$ \cite{Champion2003,Zhitomirsky2012,Savary2012}.

In two spatial dimensions, it is the Heisenberg antiferromagnet on the kagom\'e lattice that has attracted the most attention for its subtle ordering mechanisms \cite{Chalker1992,Harris1992,Ritchey1993,Huse1992,Zhitomirsky2008,Chern2013}. Its lattice geometry inherently combines stringent local constraints resulting from its short cycles (triangles) with the overall underconstrained nature due to the corner-sharing arrangement of these elementary triangles.
The quantum ground state of the spin-1/2 kagom\'e Heisenberg antiferromagnet (KHAFM) remains subject of ongoing debate,
with the formation of valence bond order contending with the emergence of different spin liquid states \cite{Yan2011,He2017a}
-- pointing to a broader scenario
where a special point is located elsewhere in the phase diagram of the KHAFM, i.e. a largely degenerate state of matter that allows
for the formation of a plethora of different states when perturbed with different interactions \cite{Hermele2008,Huh2010,Changlani2018}.

In this manuscript, we  investigate the role of chiral three-spin interactions in the context of {classical} kagom\'e antiferromagnets.
Our motivation to do so arises from the observation that for the quantum model such three-spin interactions give rise to highly unusual
states of matter: a uniform choice of chirality leads to the formation of a chiral spin liquid \cite{Bauer2014}, i.e. a bosonic analogue of a fractional quantum Hall state as first conceptualized by Kalmeyer and Laughlin \cite{Kalmeyer1987}, while a staggered pattern of chiralities leads a gapless spin liquid \cite{Bauer2019} with a spinon Fermi surface, the bosonic analogue of a metal \cite{Lee2006,Motrunich2007}.

It turns out that the (semi)classical limit of these models is very interesting as well: Akin to their quantum counterparts chiral kagom\'e antiferromagnets exhibit residual degeneracies, which in the classical systems, however, are lifted in thermal order-by-disorder transitions (observable in Monte Carlo simulations at ultralow temperatures). We also discuss interpolations to the conventional KHAFM, whose order-by-disorder phenomenology has long been established \cite{Chalker1992,Harris1992,Ritchey1993,Zhitomirsky2008,Chern2013} and in comparison to the chiral model is less restrictive in its selection of ground states.

This work is devoted to a detailed and comprehensive study of the kagom\'e antiferromagnet with chiral interactions. It is organised as follows. We first introduce the model. We then construct its ground states across the range of parameters defined by tuning the chirality from uniform to staggered, with a variable amount of admixture of the conventional Heisenberg term in the Hamiltonian. The next two chapters present the effects of thermal and quantum fluctuations, respectively, in this family of models, followed by a concluding discussion.

We find that the purely chiral models display the richest order-by-disorder phenomenology. This is largely on account of their large and elaborate ground state structure. Fluctuations can select a subset of these ground states, which at harmonic order remain exactly degenerate.

Superficially, this selection resembles the selection of coplanar `Potts' states in the pure Heisenberg antiferromagnet~\cite{Chalker1992}, but the selected chiral ground states in addition exhibit an emergent \ztwo degree of freedom. We present an effective theory for this sector of the theory, finding it well-described by a gauge-like theory of the type pioneered by Henley. This analysis culminates in the proposal of the possibility of a semiclassical route to a topological spin liquid of the toric code type upon addition of quantum dynamics.


\section{The Model}
\label{sec:model}
We consider classical spins residing on the sites of a 2D kagom\'e lattice.
The spins are 3D vectors of of unit length, $\lvert\spin{i} \rvert=1$, subject to both two- and three-body interactions.
The Hamiltonian is given by
\begin{align}
\label{eq:hamiltonian}
\mathcal{H} = &\left(1-\lvert \lambda \rvert \right) \sum_{\langle i,j \rangle}\spin{i} \cdot \spin{j} \nonumber\\
&- \lvert \lambda \rvert \sum_{i,j,k \in \bigtriangleup} \chi_{ijk} - \lambda \sum_{i,j,k \in \bigtriangledown} \chi_{ijk} \,,
\end{align}
where $\chi_{ijk}$ is the scalar spin chirality defined as
\begin{equation}
\label{eq:chirality}
\chi_{ijk}\equiv \spin{i} \cdot \left( \spin{j} \times \spin{k} \right)
\end{equation}
with $(i,j,k)$ labeling spins counterclockwise around elementary lattice triangles.
The first term in the Hamiltonian~\eqref{eq:hamiltonian} describes the usual Heisenberg interaction between nearest neighbors whereas the second and third terms describe the chiral interaction between triads of spins belonging to upright and upside-down triangles respectively.
The relative strength of these terms is parametrized by the dimensionless value $\lambda$.
We are interested in the parameter range $\lambda \in [-1,1]$.
Special cases of note are $\lambda=0$, the kagom\'e Heisenberg antiferromagnet (KHAFM), and $\lambda=\pm 1$, the uniform and staggered chiral models.
The Heisenberg interaction is antiferromagnetic everywhere inside the interval.

As a side note, we should remark that the Hamiltonian defined by Eq.~(\ref{eq:hamiltonian}) is dimensionless; its overall energy scale is set to unity. In what follows we also use $k_B=1$ and $\hbar=1$ and, consequently, temperature and frequency are also dimensionless throughout this paper.


\section{Ground state manifolds}
\label{sec:groundstates}
\subsection{The Heisenberg antiferromagnet, $\lambda= 0$}
\label{sec:heisenberg_gs}
Before turning our attention to the ground states of the general model, let us revisit the ground state manifold of the classical KHAFM.
Its Hamiltonian can be written as
\begin{equation}
\label{eq:hamiltonian_KHAFM}
\mathcal{H}_{\text{KHAFM}}= \frac{1}{2} \sum_{\bigtriangleup,\bigtriangledown} \spin{\triangle}^2 +\text{const} \,,
\end{equation}
where the sum is performed over all elementary lattice triangles with $\spin{\triangle}$ being the vector sum of the triangle's three spins.
Consequently, all ground states of the KHAFM satisfy the constraint that $\spin{\triangle}$ vanishes on each kagom\'e triangle, which is achieved by arranging all spins to form $120^\circ$ angles with their neighbors.
For each lattice triangle, such a $120^\circ$ arrangement defines a plane in which its three spins lie.
However, there is no requirement that these planes coincide for different triangles.
In other words, the ground state requirement is not sufficient to fully constrain the configuration of spins (up to some global rotation).
Consequently, the KHAFM has a large and continuous classical ground state degeneracy which can be quantified by the dimension $D$ of its ground state manifold.

\begin{figure}[t]
	\centering
	\includegraphics[width=\linewidth]{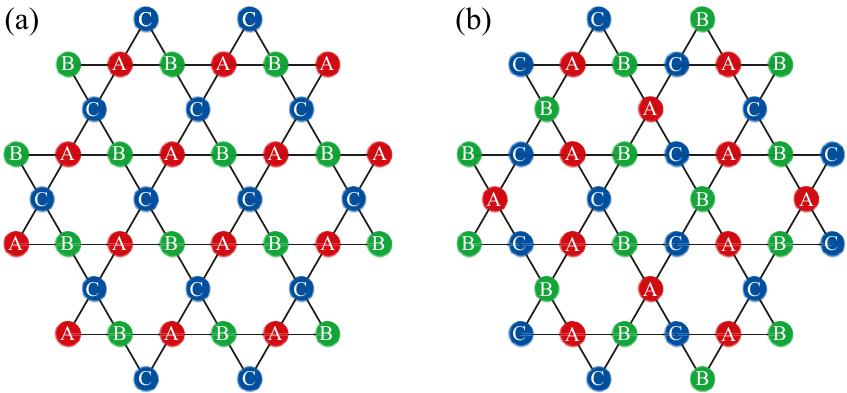}
	\caption{Two most prominent {\bf three-color states}: (a) the $q=0$ state; (b) the \rthree state. For the KHAFM the colors correspond the three possible directions of spins in a coplanar $120^\circ$ state. For the purely chiral model, A, B and C here label three possible orthogonal axes, $x$, $y$ and $z$; with the spins aligning or antialigning themselves with these axes in order to form a triad of vectors with the desired chirality/handedness on each triangle.}
	\label{fig:KHAFM_gs}
\end{figure}

Many configurations of the ground state manifold are connected by means of \textit{weathervane modes}, the energetically neutral rotation of entire chains of spins of two alternating directions around the axis defined by the spins neighboring the chain.
Crucially, a generic ground state may be obtained from a coplanar ground state (i.e., a state in which all spins lie in the same plane) through an appropriate sequence of weathervane modes~\cite{Ritchey1993,Chalker1992,Shender1996}.
Up to the global rotational symmetry, all coplanar states have spins pointing in one of three possible directions and hence can be labeled by three colors.
All distinct coplanar states are therefore in one-to-one correspondence to three-colorings of a dual honeycomb graph~\cite{Baxter1970,Huse1992} whose vertices correspond to the centers of kagom\'e triangles.
Their number is given by $Z_\text{three-color}=1.20872...^{N_{\text{vertices}}} = 1.20872...^{2N/3}$, where $N$ is the total number of spins on the kagom\'e lattice; the weathervane modes connecting different coplanar configurations correspond to exchanging two colors along any two-color loop (or a chain terminating at the boundary).
Two prominent three-color states play an important role in what follows: the $q=0$ and \rthree states shown in Fig.~\ref{fig:KHAFM_gs}.

As will be discussed later, the coplanar states play a special role in the classical KHAFM since they are selected by the so-called order-by-disorder mechanism~\cite{Chalker1992}.

\subsection{The chiral models, $\lambda = \pm 1$}
\label{sec:chiral_gs}
Scalar spin chirality $\chi_{ijk}$ is extremized when the spins are orthogonal, and its sign is determined by the handedness of the coordinate system they define.
Therefore, the ground state condition of the uniform chiral model is that the spins of all triangles form a right-handed orthogonal basis.
The ground state condition of the staggered chiral model is similar, but the spins on down-pointing triangles must instead form a left-handed orthogonal basis.

The special role played by the coplanar ground states in the KHAFM is now assumed by a subset of chiral ground states we call \textit{triaxial}.
Triaxial ground states are defined as those in which all spins are collinear with one of the three directions which define the same orthogonal basis across the entire lattice.
Like the coplanar ground states of the KHAFM, each triaxial ground state may be mapped to a three-color state on the kagom\'e lattice.
However, even after a global assignment of the three axes, this is not a one-to-one map due to the existence of an additional $\mathbb{Z}_2$ local degree of freedom that indicates whether a given spin is aligned or antialigned with a particular axis.
Therefore, in addition to a three-color assignment, the $\mathbb{Z}_2$ variables must be specified and, consequently, a triaxial ground state of the chiral model is, up to a global O(3) rotation of all spins, uniquely characterized $\mathbb{Z}_3\times\mathbb{Z}_2$ local variables. In what follows, we will borrow the terms used to describe the coplanar ground states of KHAFM such as the $q=0$ and \rthree states to refer strictly to the $\mathbb{Z}_3$, i.e. three-color arrangements; see Fig.~\ref{fig:KHAFM_gs}. Due to the existence of an additional $\mathbb{Z}_2$ degree of freedom, neither $q=0$ nor \rthree specify the actual spatial periodicity of a triaxial ground state.

Two observations are in order: (i) for any valid $\mathbb{Z}_3$ (i.e. three-color) configuration, in the case of open boundary conditions there are many distinct triaxial ground states that are different from one another by their $\mathbb{Z}_2$ variable assignments and, (ii) the number of valid $\mathbb{Z}_2$ assignments is exactly the same for all triaxial states in both uniform and staggered chiral models.
The latter statement implies that there is no statistical bias favoring some three-color configurations over others as a consequence on their additional $\mathbb{Z}_2$ multiplicity.
These properties can be demonstrated by considering a finite kagom\'e lattice of a rectangular shape.
The specific shape plays no substantive role in the argument but makes the degeneracy count straightforward.
Let the bottom row of the lattice consist of the upside-down triangles as is the case in Fig.~\ref{fig:KHAFM_gs}.
Given any three-color/triaxial arrangement, make any assignment of the $\mathbb{Z}_2$ degrees of freedom for its bottom spins.
There are $2^{L_\bigtriangledown}$ such possible assignments, with ${L_\bigtriangledown}$ being a number of upside-down triangles triangles in the bottom row.
Now assign the $\mathbb{Z}_2$ degree of freedom for the next row of spins which form the bases of these triangles.
There are $2^{L_\bigtriangledown}$ choices once again since the sign of one spin per base of a triangle can be chosen freely whereas the second spin (the last remaining spin of the triangle) is completely fixed by the triangle's chirality.
The next row of spins -- the vertices of the upright triangles -- is completely fixed but for the row after that there are $2^{L_\bigtriangledown}$ choices once again (one $\mathbb{Z}_2$ degree of freedom per base).
Proceeding this way until all sign choices have been made, we end up with $\left(2^{L_\bigtriangledown}\right)^{h_\bigtriangledown+1} = 2^{L_\bigtriangledown}\times 2^{N_\bigtriangledown}$ possible assignments regardless of the initial three-color arrangement or the signs of the individual triangles' chiralities.
(Here $h_\bigtriangledown$ is the number of rows of upside-down triangles while $N_\bigtriangledown = L_\bigtriangledown\times h_\bigtriangledown$ is the total number of such triangles, which is the same as the number of hexagons $N_{\hexagon}$ or kagom\'e unit cells.)
The prefactor of $2^{L_\bigtriangledown}$ is an artefact of the boundary condition and does not lead to any additional extensive degeneracy; in the thermodynamic limit the degeneracy of triaxial states associated with the $\mathbb{Z}_2$ degrees of freedom is 2 per unit cell of the kagom\'e lattice.

For completeness, we shall present another proof of this statement as it helps elucidate the role of boundary conditions.
Starting from any three-color state, indiscriminately assign the $\mathbb{Z}_2$ degrees of freedom for each site.
Take red to $\pm x$, green to $\pm y$, blue to $\pm z$ selecting the sign according the their $\mathbb{Z}_2$ assignments. (For simplicity, we could chose all of them to be positive -- there is nothing special about this $\mathbb{Z}_2$ assignment.)
Given the indiscriminate nature of the assignments some triangles will have the wrong handedness.
Such defects can  be ``healed'' by connecting pairs of wrong-handed triangles by a line that does not have sharp corners (i.e., its adjacent segments never make a $60^\circ$ angle) and flipping all spins situated on that line.
Requiring no sharp corners implies that the line passes through two sites of each intervening triangle, resulting in a flip of two of the triangle's spins and no change in its handedness. Meantime only one spin is flipped at the two ends of the line connecting defect triangles, which changes their handedness thus healing them.
If there are an odd number of defects, the final defect may be healed by flipping a chain which terminates at the boundary in the case of open boundary conditions.
On a compact manifold (e.g. in the case of periodic boundary conditions), an odd number of defects leave a remaining defect which may not be healed and can be associated with a $\mathbb{Z}_2$ monopole inside the surface.

This healing procedure is also a cue to the $\mathbb{Z}_2$ multiplicity count.
Specifically, two distinct  triaxial ground states with the same parent three-color state can be obtained from one another by flipping all spins along a loop containing no sharp corners.
Triangles touching these loops share exactly two spins with these loops, so the flip does not change their chirality.

Ignoring the possible effects of the boundary, the number of distinct triaxial ground states corresponding to the same three-color state is given by the number of such distinct loop configurations, which in turn is equal to the number of ways in which loops may be placed on the dual honeycomb lattice, $n_\text{triax}=2^{N_{\hexagon}}=2^{N_\bigtriangledown}$ -- the same degeneracy count that we have obtained earlier.

The aforementioned healing procedure also establishes the equivalence between classical ground states of the uniform and staggered pure chiral models.
Specifically, we can change the chirality of all upside-down triangles by flipping spins of bases of every other upright triangle in each horizontal row.

Notice, however, that the transformations between different states involving spin flips only establishes their \emph{classical} equivalence; a spin flip is not a canonical transformation for quantum spins as it does not preserve the commutation relations between their components.
As a consequence, even semiclassical features, such as spin-wave spectra discussed in Sec.~\ref{sec:weathervane_EOM} are not equivalent for the uniform and staggered pure chiral models ($\lambda=\pm 1$) as well as, generally, for different triaxial states obtained from one another by flipping all spins along a loop with no sharp corners.
The latter statement has one notable exception: as will be discussed in Sec.~\ref{sec:weathervane_EOM}, the spin-wave spectra of two triaxial states obtained from one another by flipping spins of just two out of three colors \emph{are} identical. Note that in this case flipping these spins can be achieved by means of rotating them by $\pi$ around the axis corresponding to the third color, which is a weathervane mode.

\begin{figure}[t]
	\centering
	\includegraphics[width=\columnwidth]{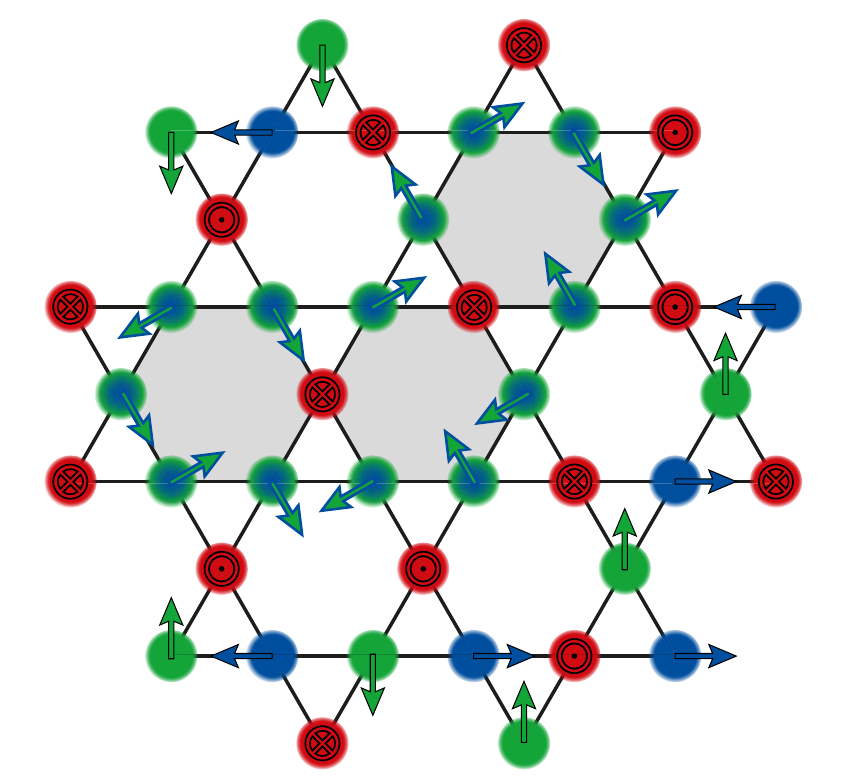}
	\caption{{\bf Weathervane modes} in a triaxial ground state of a chiral model are similar to those for the coplanar states of the KHAF: the spins aligned along two directions and forming a close loop are rotated by an arbitrary angle around the axis defined by the spins neighboring the loop. Here spins lying in the $x-y$ plane (the lattice plane) and belonging to a loop encompassing three hexagons (indicated in gray) are rotated around the $z$-axis.}
	\label{fig:weathervane_mode}
\end{figure}

In general, any triaxial state supports weathervane modes that are very similar to those of a KHAFM: all spins aligned along two alternating axes can be rotated about the third axis by the same arbitrary angle irrespective of the $\mathbb{Z}_2$ variable arrangement; an example of such a mode is shown in Figure~\ref{fig:weathervane_mode}.
Rotating these spins by $\pi/2$ results in another triaxial state whereas a rotation by $\pi$ --
a $\pi$-weathervane mode -- simply flips all spins around a loop hosting the mode.
As will be shown in Sec.~\ref{sec:weathervane_EOM} the effects of a $\pi$-weathervane rotations are mathematically equivalent to a canonical transformation of the participating spin variables which can be ``gauged out'' from the semiclassical equations of motion.

\subsection{The mixed model, $0<\lvert\lambda\rvert<1$}
\label{sec:mixed_gs}
In the mixed model, the ground state of each triangle is obtained by interpolation between three orthogonal spins and and their in-plane $120^\circ$ arrangement.
In the process the three spins ``repel'' from the space diagonal of the cube they span at $\lvert \lambda\rvert = 1$ by the same amount, until at $\lambda=0$ they end up in the plane perpendicular to that space diagonal.

\begin{figure}
	\centering
	\includegraphics[width=\linewidth]{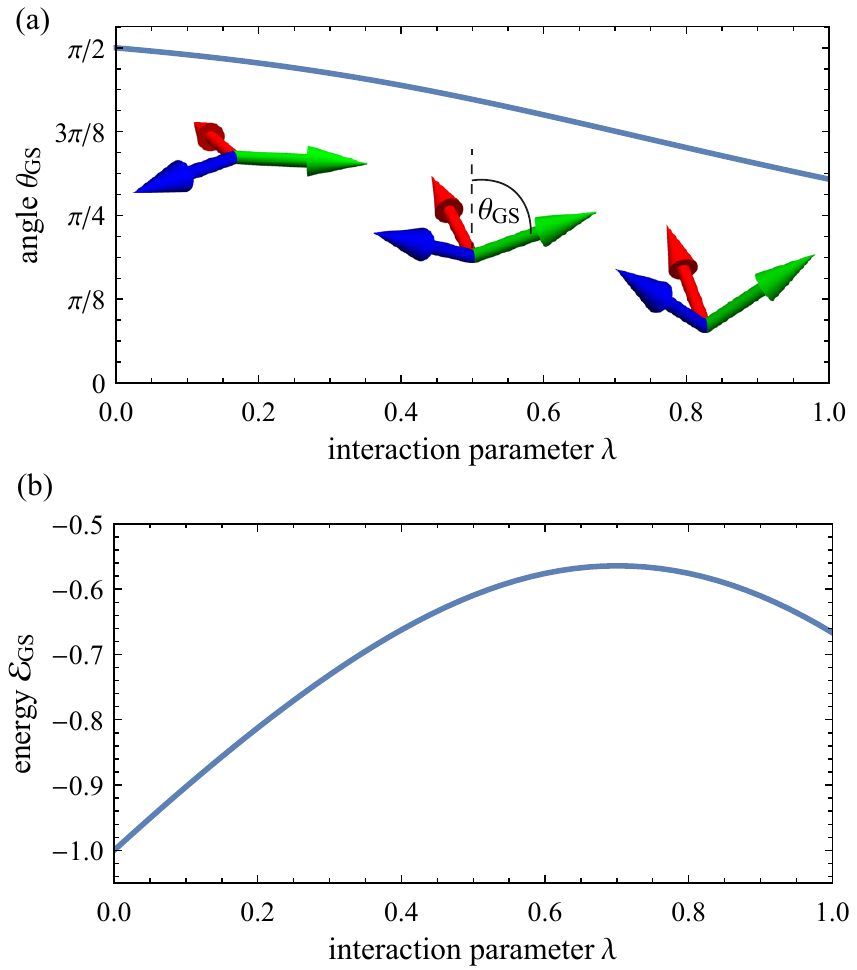}
	\caption{{\bf Triaxial ground states}. (a)~Variation of the angle between the three spin axes in the triaxial states in the mixed model, as described by Eq.~\eqref{eq:gs_angle}. (b)~Variation of the ground state energy per spin in the mixed model.}
	\label{fig:mixed_gs_energy}
\end{figure}

In terms of the angle $\theta$ that each spin makes with the space diagonal, the energy of such a configuration of three spins can be written as
\begin{equation}
\label{eq:mixed_energy}
\mathcal{E}_\triangle = \frac{3}{4}\left[ \left(1 - \lvert \lambda\rvert\right) \left(1 + 3 \cos 2 \theta\right) - \sqrt{3}\lvert \lambda\rvert\ \sin  \theta \sin 2 \theta \right] \,.
\end{equation}
An analytic expression for the angle that minimizes the energy and hence describes the ground state on each triangle is given by
\begin{equation}
\label{eq:gs_angle}
\cos \theta_\text{GS} = \frac{\lvert \lambda\rvert -1 +\sqrt{(1-\lvert \lambda\rvert)^2+\lambda^2}}{\sqrt{3}\lvert \lambda\rvert} \,.
\end{equation}
The corresponding dependence of the ground state energy on the absolute value of parameter $\lambda$ is shown in Fig.~\ref{fig:mixed_gs_energy}. The angle \emph{between} the adjacent spins in the ground state $\varphi_\text{GS}$ is given by $\cos \varphi_\text{GS} = \left(3 \cos^2 \theta_\text{GS}-1\right)/2$.

While there are multiple arrangements of spins that simultaneously minimize the energy of every triangle, satisfying the ground state condition while preserving the three-colour nature of the ground state whereby all spins are aligned along the same three directions -- see Fig.~\ref{fig:mixed_gs_energy}) -- imposes more constraints than in the pure KHAFM or pure chiral case. Specifically, in order to make three-colour ground states of individual lattice triangles compatible with one another, one needs to ensure that the space diagonals for the three spins are the same for all triangles. This is guaranteed if the three axes defined by the three colours have the chirality that minimizes the chiral terms in the Hamiltonian.  Unlike in the pure chiral case where the ``wrong'' chirality of a particular triangle can be cured by reversing an odd number of participating spins, this is no longer an option if $\lambda\neq\pm 1$ since the tree axes are not orthogonal anymore.

The states which incorporate the appropriate flux patterns are the familiar ones from the KHAFM.
For  the uniform chiral term, $0<\lambda<1$, this is the $q=0$ state, in which all arrangements of A, B, C in Fig.~\ref{fig:KHAFM_gs}a have the same chirality for all triangles. For the staggered chiral term, $-1<\lambda<0$, this is the \rthree state.

The weathervane modes associated with these three-colour ground states (i.e. with the $q=0$ state for  $0<\lambda<1$ and the \rthree state for $-1<\lambda<0$) retain the same basic nature found at $\lambda=0$ and $\lambda=\pm1$: all spins along a loop defined by two colours -- a straight line in the $q=0$ state or an elementary hexagon in the \rthree state -- can be uniformly rotated by an arbitrary angle around the axis corresponding to the third colour. However, the weathervane modes can no longer connect different three-colour states with one another. This is not at all surprising since there is a unique three-colour ground state anywhere in the mixed regime,  $\lambda\neq 0$ and $\lambda\neq\pm1$.

In addition to a unique three-colour ground state for each value of parameter $\lambda$ in the regions $-1<\lambda<0$ or $0<\lambda<1$, there are still infinitely many ground states at each $\lambda$ that are not connected to the three-colour ground states by weathervane rotations. Their construction is loosely similar to the first proof of the extensive number of distinct $\mathbb{Z}_2$ sectors associated with the triaxial states in the pure chiral case ($\lambda=\pm1$) presented in Sec.~\ref{sec:chiral_gs}; it parallels the construction of ground states in the bond-disordered KHAFM \cite{Bilitewski2017}:

Begin with the bottom row of upside-down triangles and fix the direction of the bottom spin arbitrarily for each of them. Now chose the directions of the two other spins in each of those triangles so that they satisfy the ground state condition. Since the  upside-down triangles share no spins, each of them can be satisfied independently and furthermore, there is a rotational degree of freedom associated with each of them since the top two spins can be simultaneously rotated about the direction of the fixed bottom spin with no energy cost. Now, starting from the left, use this rotational degree of freedom to arrange the correct angle, $\varphi_\text{GS}$, between adjacent spins belonging to the bases of two neighbouring upside-down triangles. This is \emph{generically} possible since for a fixed spin on the left, the spin on its right can be rotated by an arbitrary angle about the axis determined by the right spin's bottom neighbour, i.e. unrelated to the left spin. This can be visualized as follows: fix the spin on the left and define the ``target manifold'' for its right neighbour -- a circle on a unit sphere whose polar angle (measured from the left spin) is fixed to $\varphi_\text{GS}$, the angle between two adjacent spins in a ground state,  while the azimuthal angle is arbitrary. The actual possible directions of the right spin correspond to another circle with the same  polar angle but this time measured from the direction of the bottom spin, which is unrelated to the direction of the spin on the left. The two circles on the sphere would generically (albeit not always) cross at two points. (Note that they are \emph{guaranteed} to cross if $\varphi_\text{GS}=\pi$, with the choice between the two crossing points being intimately related to the $\mathbb{Z}_2$ degree of freedom for the triaxial states of the pure chiral model.) If by bad luck the circles do not cross, one can step back and try updating the choices made at previous steps until this procedure works. Once the directions of all spins in the horizontal row have been fixed, we look at their neighbours above, i.e. at the spins at the top vertices of upright triangles. Since the directions of the two spins at the base of each of those triangles are already fixed, the direction of the spin above them is uniquely determined by the ground state condition~(\ref{eq:gs_angle}). Upon this, the argument proceeds recursively until all spins have been oriented.

The overall dimensionality of the manifold associated with these ground states is subextensive in terms of continuous degrees of freedom: we have been allowed to arbitrarily orient the bottom spins and then the azimuthal angle of the left-most spin in each horizontal row is still arbitrary. There is, however, an extensive discrete degeneracy associated with the choice of two intersection points in the procedure outlined above. This choice can be generically made for each of the upside-down tringles whose number is that of the kagom\'e unit cells. Curiously, the number of weathervane modes in the \rthree state is extensive and so is the dimensionality of the GS manifold connected to the  \rthree state by weathervane rotations at $-1<\lambda<0$, with their entropy far outweighing that of the generic ground states we have just constructed. Note that the generic ground states do not have any weathervane modes, a fact that will be shown to be detrimental for their selection by the order-by-disorder mechanism.
We also remind the reader that the three-color ground states of the mixed model are unique:  the \qz ground state of the mixed model with uniform chirality and the \rthree ground state for the model with staggered chirality.


\section{Thermal order by disorder}
\label{sec:obd}

\subsection{General considerations}
\label{sec:obd:softmodes}
The low-energy spectrum of a conventional semiclassical magnetic system is described in terms of spin waves, or magnons.
Most such excitations are conventional harmonic modes.
However, one of the common features of frustrated systems is the existence of soft modes whose energy vanishes in the harmonic approximation.
At small but finite temperature, ground states with a significant number of soft modes can be selected over generic ground states by \textit{thermal order by disorder} (TObD).
Specifically, TObD occurs when harmonic contributions to the partition function from fluctuations near a subspace of the ground state manifold result in a non-integrable divergence~\cite{Moessner1998b}.
The criterion for that divergence is expressed by the inequality,
\begin{equation}
\label{eq:chalkermoessner}
D-S-M\leq0 \,,
\end{equation}
where $D$ is the dimension of the ground state manifold, $S$ is the dimension of the ground state subspace into which the system may order, and $M$ is the number of {\em soft modes} in that subspace.

The dimension of the ground state manifold $D$ may be estimated by the method  known as  Maxwell  counting, subtracting the number of ground state constraints from the number of degrees of freedom of the system
\begin{equation}
\label{eq:maxwell}
D_\text{M}=F-K \,.
\end{equation}
The Maxwell constraint counting typically underestimates the dimension since it assumes that all constraints are independent.
In cases where all constraints can be satisfied, Maxwell counting provides a lower bound, $D \geq D_\text{M}$. For that reason replacing $D$ with $D_\text{M}$ in Eq.~(\ref{eq:maxwell}) can not be used to rigorously establish the existence of TObD, nevertheless this is a useful exercise for assessing such a possibility heuristically.

Maxwell constraint counting offers the same estimate for $D$ for the entire range of $\lambda$ in the mixed chiral--Heisenberg model of Eq.~(\ref{eq:hamiltonian}).
Every spin has two degrees of freedom, and every triangle has three spins, each of which it shares with one other triangle.
That gives $F/N_{\bigtriangleup,\bigtriangledown}=2 \times 3/2=3$ degrees of freedom per triangle, with $N_{\bigtriangleup,\bigtriangledown}$ being the number of triangles.
Irrespective of $\lambda$, the number of constraints is also $K/N_{\bigtriangleup,\bigtriangledown}=3$: The direction of the first spin is arbitrary (no constraints), the second spin has a fixed polar angle (one constraint) but unconstrained azimuth, the direction of the third spin is completely fixed (two more constraints).
An alternate perspective is this: Each triad of spins has six degrees of freedom, but once the ground state condition has been met for a particular triangle (see Fig.~\ref{fig:mixed_gs_energy}), only three remain, the three Euler angles.
Therefore the number of constraints is again $K/N_{\bigtriangleup,\bigtriangledown}=6-3=3$.
Consequently, just like the KHAFM, both mixed and pure chiral models are marginally constrained according to Maxwell counting: $D_\text{M}=0$.

A tighter lower bound on $D$  can be obtained by considering the independent ``directions'' by which one may leave a particular GS state while remaining in the GS manifold.
For the \rthree state, there is one independent weathervane mode for every unit cell  i.e., one for every three unit cells of the kagom\'e lattice itself.
The dimension of the GS manifold must be at least this large, provided that the \rthree state is one of the ground states, which is the case for $-1\leq\lambda\leq 0$ and $\lambda = 1$.
This observation is crucial because it demonstrates that $D\geq N{\bigtriangledown}/3$ (where $N_{\bigtriangledown}=N/3$ is the  number of upside-down triangles, which is equal to that of unit cells of the kagom\'e lattice) is extensive in system size.
Therefore, whenever the \rthree state is one of the ground states, i.e. for $-1\leq\lambda\leq 0$ and $\lambda  = 1$, the dimensionality of the ground state manifold $D\geq N_{\bigtriangledown}/3$, and we expect this bound to be tight since the \rthree state has the largest number of weathervane modes.

For the mixed model with uniform chirality, $0<\lambda<1$, the \rthree state is no longer a ground state and therefore this bound does not apply.
Instead, the ground state with the greatest number of independent weathervane modes is the \qz state. Since all weathervane modes must be completely straight in this case,
the corresponding lower bound is proportional to the perimeter of the system rather than the area, i.e. $D$ is no longer extensive.

Note that $S$ in Eq.~(\ref{eq:chalkermoessner}) is never extensive and hence the possibility
of TObD hinges on the number of soft modes $M$. To avoid confusion, we should reiterate that the soft modes that we need to consider here are \emph{not} the same as weathervane modes: any mode with sub-harmonic spectrum counts as soft for the purposes of the criterion given by Eq.~(\ref{eq:chalkermoessner}).

Just like in the case of KHAFM~\cite{Chalker1992}, the modes of interest turn out to be spin waves with \emph{quartic} dispersion; in what follows we will evaluate their number.
Notably, such soft modes exhibit a distinct signature in the low-temperature specific heat: Whereas harmonic modes contribute $k_B/2$ to the specific heat, quartic soft modes contribute only $k_B/4$.
As a consequence, the onset of TObD manifests itself through a diminished specific heat as the temperature is lowered.

\subsection{The Heisenberg antiferromagnet, $\lambda= 0$}
\label{sec:ToBD_KHAFM}

As a primer, let us put the above statements into context for the case of the pure KHAFM ($\lambda=0$) first.
Here all coplanar ground states have been shown to have the same harmonic-order spectrum~\cite{Chalker1992}.
Furthermore, there is exactly one such soft mode per unit cell of the kagom\'e lattice.
The remaining five are conventional harmonic modes.
The TObD criterion in Eq.~(\ref{eq:chalkermoessner}) thus gives (ignoring subextensive $S$) $D-M\sim N_{uc}/3-N_{uc}<0$, strongly suggesting TObD ordering into coplanar states. Further selection within the coplanar state manifold then takes place via anharmonic fluctuations~\cite{Huse1992}, in favor of a weak \rthree order parameter~\cite{Chern2013}. Due to the rather special feature of the KHAFM -- the independence of the spin wave spectrum of the details of an underlying coplanar ground state -- such discrimination between different coplanar states cannot be argued at the harmonic level. This special property does not survive in the presence of the chiral interaction as we will show in the following section.

The selection of the coplanar states in the KHAFM is marked by a
specific heat per lattice unit cell (with 5 quadratic modes and one soft mode per unit cell) of $(5\times1/2 + 1\times1/4) k_B = 11/4 k_B$ or a specific heat per spin,
\begin{equation}
\label{eq:eleventwelvths}
C/N= \frac{11}{12}\, k_B \,,
\end{equation}
a signature of TObD observed in Monte Carlo simulations~\cite{Chalker1992}.

\subsection{The chiral models, $\lambda = \pm 1$}
\label{sec:low_energy}

\begin{figure}[b]
	\centering
	\includegraphics[width=\linewidth]{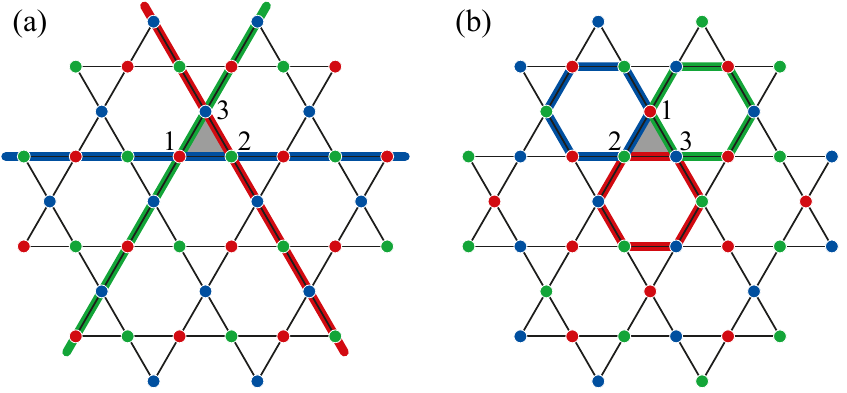}
	\caption{{\bf Interaction of weathervane modes} in a ground state order of type (a) \qz and (b) \rthree. The three-coloring of sites indicates the orientation of spins pointing in either one of the three orthonormal directions allowed at $\lambda=\pm 1$. The colored lines identify weathervane modes -- i.e., collections of spins which can be rotated around a common axis (with the axis color coded in analogy to the three spin orientations) at zero energy cost, since they only couple to outside spins which are parallel to the rotation axis. The crossing of two weathervane modes, which occurs at sites labeled $1$, $2$, and $3$, is associated with an energy cost that is quartic in the rotation angles (see text for details). Every triangle is affected at most by three weathervane modes, as shown here for the gray shaded triangles. }
	\label{fig:weathervanemodes}
\end{figure}

Unlike coplanar ground states, different triaxial states do not have the same harmonic spectrum.
Instead, the soft modes are now associated with attempting to implement two ``intersecting'' weathervane modes, i.e. weathervane modes sharing a spin. Since the spin at the intersection cannot simultaneously participate in two incompatible modes, this results in a ``kink'' that lifts the ground state into an excited state.
As shown below, in the case of the pure chiral models, the minimum energy cost of this kink is quartic in the amplitudes (i.e. rotation angles) of the two participating weathervane modes.

Without loss of generality we demonstrate this on the example of the \qz and \rthree configurations.
Note that the specific type of order does not make a difference for our considerations: As shown in Fig.~\ref{fig:weathervanemodes}, every triangle in the kagom\'e lattice is affected by at most three weathervane modes; every spin in the lattice is affected by at most two weathervane modes, irrespective of the specific ground state order.
Our goal is to compute the local energy change on such a maximally-affected triangle as a function of the rotation angles.

For simplicity, we assume that the three spins $\mathbf{S}_1$, $\mathbf{S}_2$, and $\mathbf{S}_3$ on the triangle point along the $x$, $y$, and $z$ direction, respectively, forming an orthonormal basis and maximizing the chirality $\chi_{123}=1$.
We introduce a weathervane rotation $R^x_\alpha$ around the $x$-axis by an angle $\alpha$, which affects the spins $\mathbf{S}_2$, and $\mathbf{S}_3$ (see Fig.~\ref{fig:weathervanemodes}), transforming them into $R^x_\alpha\mathbf{S}_2$ and $R^x_\alpha\mathbf{S}_3$, respectively.
Similarly, the subsequent introduction of a second weathervane rotation around the $y$-axis by an angle $\beta$ affects spins $\mathbf{S}_1$, and $\mathbf{S}_3$.
The third rotation around the $z$-axis by an angle $\gamma$ affects spins $\mathbf{S}_2$, and $\mathbf{S}_3$, such that we obtain the transformed set of spins
\begin{align}
\mathbf{S}_1' &= R^z_\gamma R^y_\beta \mathbf{S}_1 \nonumber\\
\mathbf{S}_2' &= R^z_\gamma R^x_\alpha \mathbf{S}_2 \nonumber\\
\mathbf{S}_3' &= R^y_\beta R^x_\alpha \mathbf{S}_3 \,.
\end{align}
The chirality $\chi_{123}'$ on the triangle after performing the spin rotations can now be computed and the result expanded in the rotation angles to obtain
\begin{equation}
\chi_{123}'=1-\frac{\alpha^2\beta^2}{2} - \frac{\alpha^2\gamma^2}{2} - \frac{\beta^2\gamma^2}{2} + \mbox{sixth-order terms} \,.
\end{equation}
We thus obtain three independent quartic modes at the intersection of three weathervane defects.
The intersection of two weathervane defects, i.e.\ setting one of the three angles to zero, would imply the emergence of a single quartic mode.
With this insight, we can make the central observation: Upon introducing an increasing number of weathervane defects to the system, the total number of independent quartic modes equals the number of independent rotation angles in the system -- which is equivalent to the number of weathervane defects present.

\subsubsection{Soft modes of the \qz state}
\label{sec:soft_modes_q0}
The weathervane modes of the \qz state wrap around the torus.
The number of soft modes is therefore linear in the circumferences of the torus, and scales as the square root of the area.
Since there are three species of weathervane modes, the number of soft modes is roughly three times the circumference.
This falls short of the lower-bound on the dimension of the ground state manifold obtained from the \rthree state, which scales linearly with area.
Therefore, we do not expect to observe \qz order of the pure chiral models.

\subsubsection{Soft modes of the \rthree state}
\label{sec:soft_modes_root3}
On the other hand, the weathervane modes of the \rthree states are localized to single hexagons.
Their number is equal to the number of kagom\'e unit cells.
Hence, they are equal in number to the soft modes of the KHAFM coplanar states.
They outnumber our lower bound on the dimension $D$ by a factor of three.
Since the \rthree states are discrete, they have no continuous ``dimension'' aside from the overall rotational symmetry of the entire state.
Therefore $S$ plays no role in the thermodynamic limit.

\subsubsection{A remark on the magnon spectra: flat and grooved bands}
\label{sec:flat_and_grooved}
Having identified the soft modes of these two classes of ground states, we can immediately remark on some of the characteristics of their lowest magnon bands in linear spin-wave theory.
Since the soft modes of the \qz states scale as the length of the system, they are capable of supporting lines of zero-energy in the lowest magnon band: flat grooves run through the Brillouin zone.
Being sub-extensive, however, there are insufficient soft modes to form a complete flat band.
The direction of these zero-grooves may also be inferred from the states.
These lines also exist in the exact spectra of magnons since implementation of non-intersecting weathervane modes may be set up for any wavevector which is perpendicular to the weathervane modes.
In other words, the weathervane modes may be seen as wavefronts of exact zero-energy fluctuations.

On the other hand, the soft modes of the \rthree class of states are extensive in system size.
They scale with the area of the system, and are of sufficient number to form a completely flat band (to harmonic order).
Of course these flat bands are exactly flat too, since the non-intersecting hexagonal weathervane modes support exactly zero-energy waves with any wavevector in any direction whatsoever.

We come back to a more detailed discussion of the magnon spectra of the \qz and \rthree states in Sec.~\ref{sec:uniform_spectra} below.

\subsection{The mixed model, $0<\lvert\lambda\rvert<1$}
\label{sec:low_energy_mixed}

\subsubsection{Soft modes of the mixed models}
\label{sec:soft_modes_mixed}
Soft modes also exist for intermediate values of $\lambda$, but there is no purely entropic competition between the \qz and \rthree states since they never both meet the ground state condition for non-integer values of $\lambda$.
Nonetheless, they are each favored in the domains in which they are ground states since the soft modes are counted by the total number of weathervane modes, but our lower limit on the dimension of the ground state manifold is obtained by the number of \textit{independent} weathervane modes.
For both positive and negative non-integer values of $\lambda$, the number of soft modes, $M$, exceeds our lower limit on the dimension of the ground state manifold, $D$ by a factor of three.
Therefore, selection of the \rthree state is expected for $-1<\lambda<0$, and selection of the \qz state is expected for $0<\lambda<1$.

\subsubsection{Monte Carlo simulations}
\label{sec:monte_carlo}

Throughout the discussion of the ground state manifold of the mixed model in Sec.~\ref{sec:mixed_gs}, we stated that different order-by-disorder mechanisms are expected to occur depending on the sign of the interaction parameter $\lambda$.
In order to resolve the thermal selection of distinct ground state configurations, we perform classical Monte Carlo simulations of the model Hamiltonian Eq.~\eqref{eq:hamiltonian} in the full parameter range $\lambda\in\left[ -1,1 \right]$, i.e., covering the mixed model as well as the limiting cases of the pure chiral models and the Heisenberg antiferromagnet.

The simulations are performed on a finite lattice with periodic boundary conditions, comprising $L\times L$ unit cells at $L=6$ and a total of $N=108$ spins.
In order to resolve even lowest temperatures, we employ a parallel tempering scheme for the concurrent simulation of 136 logarithmically spaced temperature points between $T_\mathrm{min}=10^{-5}$ and $T_\mathrm{max}=10$.
Within every simulation, we prepare the system in a coplanar \qz state and let it thermalize for $10^9$ sweeps, each attempting $N$ random local updates of the spin configuration, before taking measurements for another $10^8$ sweeps.

\begin{figure}[b]
\centering
\includegraphics[width=\linewidth]{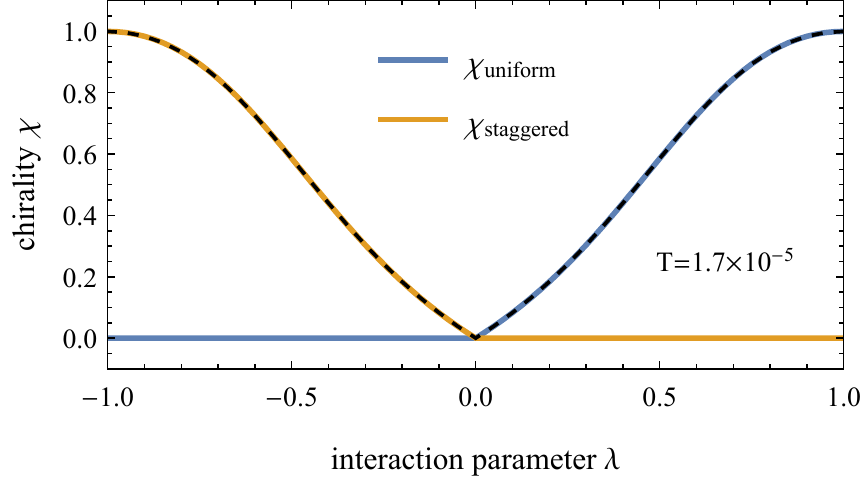}
\caption{\textbf{Chiral order parameter} of the mixed chiral-Heisenberg model at $T=10^{-5}$. For $\lambda<0$ the staggered chirality becomes finite while the uniform chirality remains zero. Vice versa, at $\lambda>0$ the uniform chirality assumes finite values. In both cases the values exactly coincide with the chiral-Heisenberg model on a single triangle (dashed line) where a continuous flattening of orthogonal triaxial states into coplanar triaxial states is expected (see Fig.~\ref{fig:mixed_gs_energy}).}
\label{fig:mc:chirality}
\end{figure}

We begin our study with an analysis of the scalar spin chirality, as defined in Eq.~\eqref{eq:chirality}.
In order to discriminate the uniform chiral model and the staggered chiral model, we define the uniform chiral order parameter
\begin{equation}
\chi_\mathrm{uniform} = \frac{1}{N_\bigtriangleup + N_\bigtriangledown}\sum\limits_{ijk \in \bigtriangleup,\bigtriangledown} \chi_{ijk} \,,
\end{equation}
which computes the average chirality on every triangle and is therefore maximized when the chirality has the same sign on all triangles,
and the staggered chiral order parameter
\begin{equation}
\chi_\mathrm{staggered} = \frac{1}{N_\bigtriangleup}\sum\limits_{ijk \in \bigtriangleup} \chi_{ijk} - \frac{1}{N_\bigtriangledown}\sum\limits_{ijk \in \bigtriangledown} \chi_{ijk} \,,
\end{equation}
which is maximized when the chirality has opposite signs on up-pointing and down-pointing triangles.
The normalization constants $N_\bigtriangleup$ and $N_\bigtriangledown$ denote the total numbers of upright and upside-down triangles in the lattice.
Calculating these order parameters, we confirm that the uniform chiral order parameter becomes finite for $\lambda>0$ and the staggered chiral order parameter is finite for $\lambda<0$. Moreover, the magnitude of the chirality coincides with the value that is expected from triaxial states at canting angles as defined in Eq.~\eqref{eq:mixed_energy}, see Fig.~\ref{fig:mc:chirality}.

\begin{figure*}
\centering
\includegraphics[width=\linewidth]{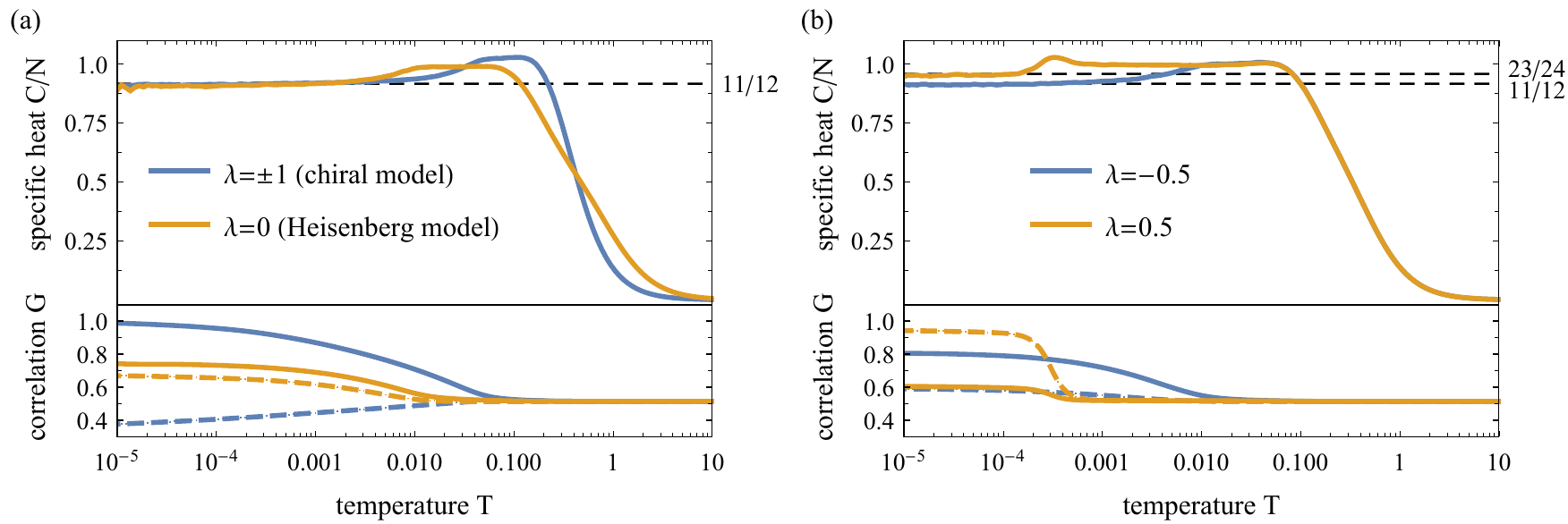}
\caption{\textbf{Specific heat and spin correlations} as a function of temperature. (a) Specific heat in the limiting cases of purely staggered chiral interactions ($\lambda=-1$), uniform chiral interactions ($\lambda=1$), and Heisenberg interactions ($\lambda=0$), respectively. The curves of the two chiral models exactly coincide. In all three cases the specific heat approaches $C/N=\frac{11}{12}$ at low temperatures as expected from the counting of soft modes. The spin correlations $G_{\sqrt{3}\times\!\sqrt{3}}$ (solid curves) build up around the same temperature scale where the specific heat approaches its low-temperature value and dominate over the $G_{q=0}$ correlations (dashed curves).
(b) Competition between energetic constraints on the ground state and entropic corrections (which is present at $\lambda=0.5$ but not at $\lambda=-0.5$, see text for details) leads to a strongly suppressed ordering temperature.
The specific heat converges to $C/N=\frac{23}{24}$ ($\frac{11}{12}$), as expected for \qz (\rthree) order on a finite lattice with $L=6$ (see text for details). Likewise, the spin correlations $G_{q=0}$ at $\lambda=0.5$ build up at a lower temperature than the spin correlations $G_{\sqrt{3}\times\!\sqrt{3}}$ for $\lambda=-0.5$.  }
\label{fig:mc:heatTemperature}
\end{figure*}
Within our simulation of the thermodynamic properties of the system, it is our prime goal to study the onset of magnetic order -- either of the \rthree state or of \qz type -- as a consequence of thermal fluctuations.
Signatures thereof are extracted from the specific heat.
Following the counting argument for the contribution of soft modes outlined in Sec.~\ref{sec:ToBD_KHAFM}, we expect the specific heat per spin to approach $C/N=11/12$ for the \rthree state and $C/N=1$ for the \qz state in the thermodynamic limit.
On finite lattices, in contrast, we may expect finite-size corrections.
The number of soft modes in the \rthree state equals the number of hexagons in the lattice, which is $N_\mathrm{soft}=L^2$; in the \qz state, the number of soft modes equals the number of straight lines spanning the lattice, which scales sub-extensively as $N_\mathrm{soft}=3L$.
Attributing a specific heat of $1/4$ to every soft mode and a contribution to $1/2$ to each of the remaining $2N-N_\mathrm{soft}$ quadratic modes, the specific heat of the finite $L=6$ \rthree state amounts to
\[
	C/N=11/12 \,,  ,
\]
precisely matching its value in the thermodynamic limit.
In contrast, the specific heat of a {\em finite} $L=6$ \qz configuration is
\[
	C/N=23/24 \,,
\]
while it would approach $1$ in the thermodynamic limit.

For the Heisenberg model as well as for the chiral models we observe a drop of the specific heat to $11/12$ at temperatures below $T\approx 10^{-2}$, suggesting the onset of \rthree order.
This behavior also persists in the mixed model at $\lambda=-0.5$, i.e., in the case of staggered chiralities. At $\lambda=0.5$, however, the specific heat approaches a value of $23/24$, which is compatible with \qz order, see Fig.~\ref{fig:mc:heatTemperature}.
We note that the transition into the \qz state is difficult to resolve within Monte Carlo simulations with only local updates, since it is driven by the existence of soft modes that involve the rotation of spin chains spanning the entire system. Consequently, when cooling down from a random configuration, the simulations fail to equilibrate into the \qz ordered state (in which case the specific heat remains unity, indicating the absence of any soft modes), which is why we chose to initially prepare the system in a coplanar \qz configuration instead (that is still different from the non-coplanar, triaxial \qz ground state).

\begin{figure}
\centering
\includegraphics[width=\linewidth]{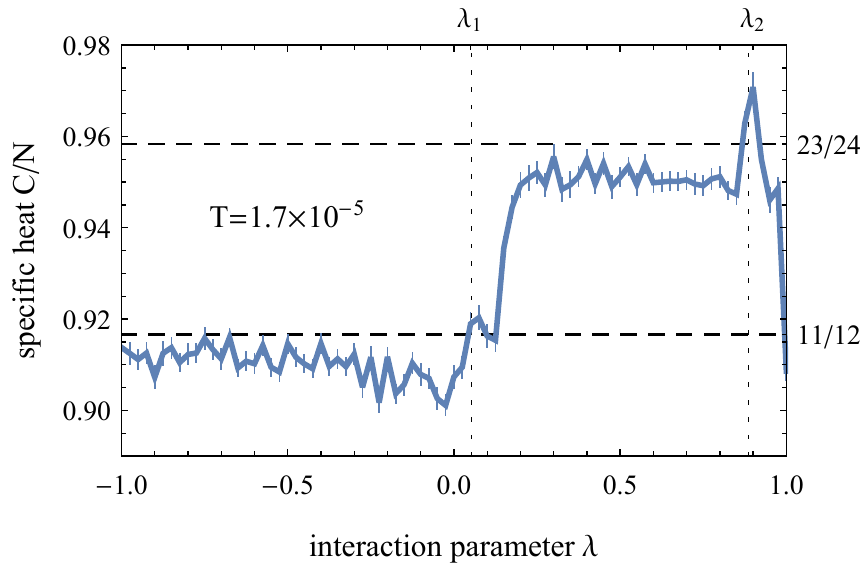}
\caption{\textbf{Specific heat} as a function of the interaction parameter $\lambda$ in the low-temperature regime $T=1.7\cdot 10^{-5}$. The specific heat per spin assumes values of either $C/N=11/12$, associated with the \rthree state or values of $C/N=23/24$, associated with the \qz state (see text for details). The dotted lines indicate phase boundaries as determined by the spin correlations shown in Fig.~\ref{fig:mc:op}.
}
\label{fig:mc:heat}
\end{figure}

In regimes where \qz order prevails we also observe a strong suppression of the transition temperature, with the specific heat assuming its final low-temperature value only below $T\approx 10^{-4}$.
The above observations are in line with our previous claim that the \rthree configuration is not a ground state of the mixed model with uniform chiralities, but nevertheless it competes with the \qz configuration due to its larger number of soft modes, contributing favorably to the free energy.
The competition between the two states becomes relevant as $\lambda$ approaches zero or unity, when the energy splitting between the \qz ground state and the \rthree configuration becomes small.
A systematic analysis of the low-temperature specific heat as a function of the interaction parameter confirms that it is compatible with the \rthree for all values $\lambda \leq 0$. It assumes values that are compatible with \qz order for a large portion of the $\lambda>0$ regime; A peak in the specific heat at $\lambda_2 \approx 0.9$ signals the phase boundary beyond which the \rthree ground state of the uniform chiral model becomes more favorable, see Fig.~\ref{fig:mc:heat}.
The lower phase boundary $\lambda_1 \approx 0.05$ is less distinct in our calculations of the specific heat, but it is better visible in spin correlation functions.

We therefore corroborate our findings about the order-by-disorder driven phase transitions by investigating spin correlation functions which serve as order parameters of the two potential ground state configurations.
For this purpose, we define the sublattice correlation function
\begin{equation}
G(L_k) = \sum\limits_{i,j \in L_k} \left| \mathbf{S}_i \cdot \mathbf{S}_j \right| \,,
\end{equation}
which determines the absolute value of spin correlations within a given sublattice $L_k$.
Partitioning the kagom\'e lattice into its three sublattices $L_1^{q=0}$, $L_2^{q=0}$, and $L_3^{q=0}$ which constitute the \qz configuration (c.f.~Fig.~\ref{fig:KHAFM_gs}), we define the global \qz correlation function
\begin{equation}
G_{q=0} = \frac{1}{N} \sum\limits_{k=1}^3 G\left( L_k^{q=0}  \right) \,,
\end{equation}
which is expected to approach unity if the system assumes long range order of \qz type.
Similarly, we define the global correlation function
\begin{equation}
G_{\sqrt{3}\times\!\sqrt{3}} = \frac{1}{N} \sum\limits_{k=1}^3 G\left( L_k^{\sqrt{3}\times\!\sqrt{3}} \right)
\end{equation}
to identify long-range order of \rthree type.
We point out that, by taking the absolute value of the spin correlations, we deviate from conventional definitions.
However, by virtue of this construction, the correlation function becomes insensitive to the sign of the involved spins, thereby extending its applicability to also identify triaxial configurations with additional $\mathbb{Z}_2$ degrees of freedom.
At the same time, our definition of the correlation functions implies that they become non-zero when applied to a configuration that dos not match their sublattice structure (e.g. $G_{q=0}$ applied to a \rthree configuration).
The exact lower value depends on the canting angle of the triaxial state defined in Eq.~\eqref{eq:gs_angle}.

We observe that the appropriate correlation function becomes large in the low-temperature regime, with the correlations building up around the same temperature scale where the specific heat transitions into its low-temperature value, see Fig.~\ref{fig:mc:heatTemperature}.
In contrast to the Heisenberg antiferromagnet, where the \rthree correlations become dominant but do not saturate, in accordance to observations in previous studies~\cite{Chern2013}, the same correlation function is maximized by the chiral models at temperatures as low as $T=10^{-5}$.
For the mixed model, particularly at $\lambda>0$, the spin correlations reflect a suppressed ordering scale that is in agreement with our specific heat measurement.
Systematically mapping out the dependence of the spin correlations on the interaction parameter in the low-temperature regime allows us to distinguish three different regimes, as shown in Fig.~\ref{fig:mc:op}:
\begin{figure}
\centering
\includegraphics[width=\linewidth]{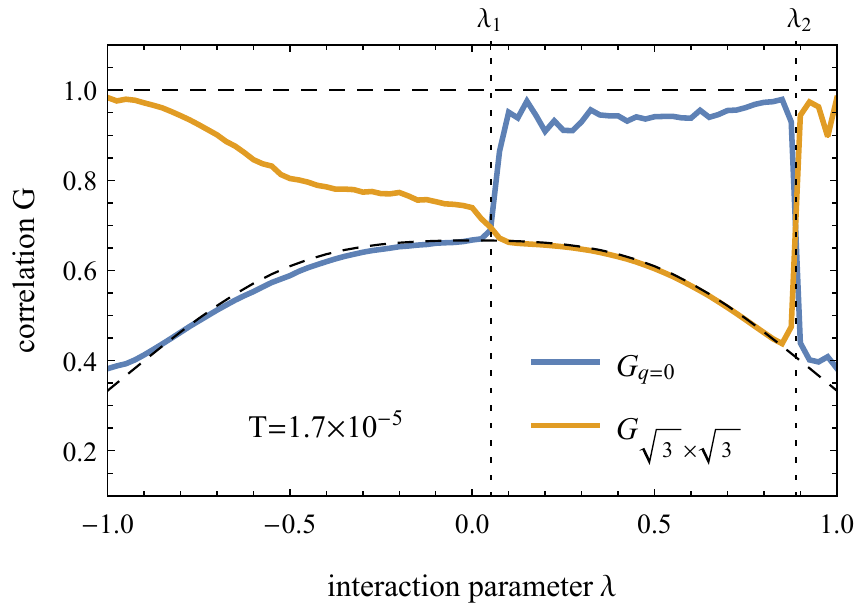}
\caption{\textbf{Spin correlation functions} plotted as a function of the interaction parameter $\lambda$ in the low-temperature regime $T=1.7\cdot 10^{-5}$. The \qz correlation function calculated for a \qz state is unity (top dashed line). Applying the \qz correlation function to a \rthree configuration, and vice versa, returns a smaller value (lower dashed line) which depends on the canting angle defined in Eq.~\eqref{eq:gs_angle}. The entropically driven phase transitions at $\lambda_1\approx 0.05$ and $\lambda_2\approx 0.9$ are indicated by the dotted lines. }
\label{fig:mc:op}
\end{figure}
(i) A region of \rthree order extending from $\lambda=-1$ to $\lambda_1\approx 0.05$.
The upper phase boundary extends beyond the Heisenberg point into the domain where the \rthree configuration is stabilized only entropically, while not being a true ground state of the system.
Consequently, the location of the phase transition depends on the temperature and $\lambda_1$ shifts towards larger values when the temperature is increased.
(ii) A region between $\lambda_1$ and $\lambda_2\approx 0.9$ where the system assumes its \qz ground state configuration.
(iii) A region of \rthree order between $\lambda_2$ and $\lambda=1$ where, similar to the phase transition near the Heisenberg point, the magnetic order is stabilized for a small region away from the uniform chiral point despite not being a ground state.
The transition point $\lambda_2$ shifts towards smaller values when the temperature is increased, corroborating that the selection of the \rthree state in the regime $\lambda \geq \lambda_2$ is through an entropic mechanism, i.e.\ via TObD. In particular, the entropic contribution to the free energy outweighs the energetic preference for the \qz state for $\lambda<1$.
We dub this a ``proximate order-by-disorder" regime in analogy to the notion of a proximate spin liquid \cite{Banerjee2016,Banerjee2017,Revelli2020} in systems where thermal fluctuations expand the extent of a quantum spin liquid to the finite-temperature regime above a nearby ordered ground state.

\section{Quantum order by disorder}
\label{sec:QOBD}
From a semiclassical perspective, the degeneracy among classical ground states can  also be lifted by quantum fluctuations via a quantum order-by-disorder mechanism. Practically this means that the role of the free energy in selecting between different states at a finite temperature is taken at $T=0$ by the zero-point energy associated with spin excitations.
The magnon spectrum in various classical ground states of relevance here can be obtained through Fourier transforming and then numerically diagonalizing coupled semiclassical equations of motion for the spin degrees of freedom. Since each (positive) magnon frequency $\omega(\mathbf{k})$ contributes $\omega(\mathbf{k})/2$ to the total zero-point energy ($\hbar=1$), integration of $\omega(\mathbf{k})/2$ over the Brillouin zone and summation over bands yields the quantum harmonic correction to the energy of a classical ground state. Quantum order by disorder (QObD) then refers to the mechanism whereby a system selects a state with the lowest quantum corrections to its classical ground state energy. Crudely speaking, the states with the softest
modes win.

For the system at hand, the additional $\mathbb{Z}_2$ degree of freedom that hugely increases the degeneracy of classical triaxial ground states of the pure chiral model ($\lambda=\pm 1$) in comparison to the KHAFM ($\lambda=0$) case presents an additional complication; it is clearly unfeasible to calculate quantum corrections to the ground-state energy for all $\mathbb{Z}_2$ sectors within the same triaxial state. However, as we demonstrate below in Sec.~\ref{sec:weathervane_EOM}, the classical ground states that are only different by $\pi$-weathervane rotations have exactly the same harmonic correction to their energy, so that not all sectors need to be computed separately. The remaining number of  $\mathbb{Z}_2$ sectors to be computed, however, still remains large, and to avoid having to do an explicit calculation for each of these, we devise an effective theory outlined in Sec.~\ref{sec:effective}.

\subsection{Semiclassical equations of motion in the chiral model}
\label{sec:EOM}

To set the stage, we first derive the semiclassical equations of motion (EOMs) of spins. These can be used to study the quantum fluctuations in the classical ground states of the pure chiral Hamiltonian ($\lambda=\pm 1$) in the harmonic approximation.

We start by noting that whenever all terms of a spin Hamiltonian that involve a particular spin $\mathbf{S}_i$ can be written in a form
\begin{equation}
\mathcal{H}_{i}
= - \mathbf{\tilde{h}}_i\cdot\mathbf{S}_i,
\label{eq:hamiltonian_part}
\end{equation}
where $\mathbf{\tilde{h}}_i$ is an effective Zeeman field that depends on some subset of the remaining spins, the semiclassical EOM is simply given by
\begin{equation}
\label{eq:Landau_Lifshitz}
\frac{d \mathbf{S}_i}{dt}=\mathbf{S}_i\times\mathbf{\tilde{h}}_i,
\end{equation}
the so-called Landau--Lifshitz equation.  (Note that the entire Hamiltonian $\mathcal{H}\neq \sum_i \mathcal{H}_i$ due to multiple counting of the spin interaction terms.)

\begin{figure}[h!]
	\centering
	\includegraphics[width=0.37\columnwidth]{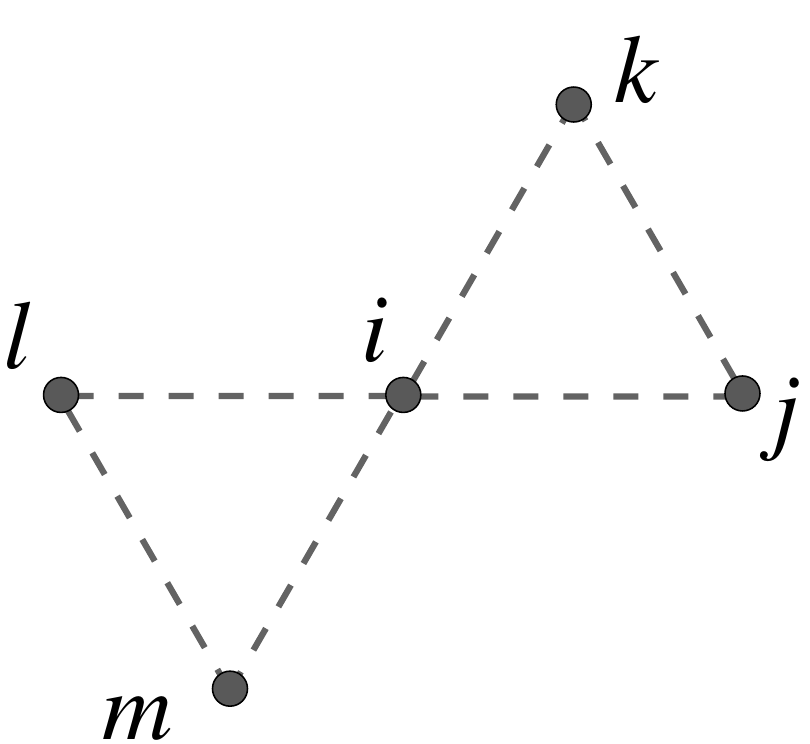}
	\caption{Labelling of sites of a kagom\'{e} cluster used in Section~\ref{sec:weathervane_EOM}.}
	\label{fig:cluster_labels}
\end{figure}

Let us apply this to the specific Hamiltonian of Eq.~(\ref{eq:hamiltonian}) with $\lambda=\pm 1$, i.e. on the pure chiral model. Here we have
\begin{equation}\label{eq:effective_field}
 \mathbf{\tilde{h}}_i = \left( \mathbf{S}_j \times \mathbf{S}_k\right)
\pm \left( \mathbf{S}_l \times \mathbf{S}_m\right)
\end{equation}
with the spins labelled as shown in Fig.~\ref{fig:cluster_labels}.
Therefore the EOM turns into
\begin{multline}
\label{eq:chiral_EOM}
\dot{\mathbf{S}}_i
= \mathbf{S}_j \left(\mathbf{S}_i\cdot \mathbf{S}_k\right) - \mathbf{S}_k \left(\mathbf{S}_i\cdot \mathbf{S}_j\right)\\
\pm \mathbf{S}_l \left(\mathbf{S}_i\cdot \mathbf{S}_m\right) \mp \mathbf{S}_m \left(\mathbf{S}_i\cdot \mathbf{S}_l\right).
\end{multline}
Since in any of the ground states of the chiral model the neighboring spins are perpendicular to one another, 
linear contributions to the right-hand side of Eq.~(\ref{eq:chiral_EOM}) can come solely from the dot products. Consequently, the linearized EOM for $\mathbf{S}_i$ in a particular ground state reads
\begin{multline}
\label{eq:chiral_EOM_lin}
 \frac{d\,\delta\mathbf{S}_i}{dt}
= \Big[\overline{\mathbf{S}}_j \left(\delta\mathbf{S}_i\cdot \overline{\mathbf{S}}_k + \delta\mathbf{S}_k\cdot \overline{\mathbf{S}}_i\right)\\
-\overline{\mathbf{S}}_k \left(\delta\mathbf{S}_i\cdot \overline{\mathbf{S}}_j + \delta\mathbf{S}_j\cdot \overline{\mathbf{S}}_i\right)\Big]\\
\pm \Big[j\to l; k\to m\Big],
\end{multline}
where ${\mathbf{S}}_n=\overline{\mathbf{S}}_n + \delta{\mathbf{S}}_n$ with $\overline{\mathbf{S}}_n$  denoting a spin in the ground state configuration.
Due to the fact that $\dot{\mathbf{S}}_i\perp \overline{\mathbf{S}}_i$ and the mutual orthogonality of all neighboring spins in their ground state configuration, the ``toy'' EOMs for the two components of $\dot{\mathbf{S}}_i$ read
\begin{subequations}
  \label{eq:EOM_scalar}
  \begin{multline}
      \label{eq:EOM_scalar_a}
    \frac{d\left(\delta\mathbf{S}_i\cdot \overline{\mathbf{S}}_j\right)}{dt}
    = \delta\mathbf{S}_i\cdot \overline{\mathbf{S}}_k + \delta\mathbf{S}_k\cdot \overline{\mathbf{S}}_i\\
    \pm \Big[\overline{\mathbf{S}}_j\cdot\overline{\mathbf{S}}_l\ \left(\delta\mathbf{S}_i\cdot \overline{\mathbf{S}}_m + \delta\mathbf{S}_m\cdot \overline{\mathbf{S}}_i\right)\\
    -\overline{\mathbf{S}}_j\cdot\overline{\mathbf{S}}_m \left(\delta\mathbf{S}_i\cdot \overline{\mathbf{S}}_l + \delta\mathbf{S}_l\cdot \overline{\mathbf{S}}_i\right)\Big],
\end{multline}
    \begin{multline}
    \label{eq:EOM_scalar_b}
    \frac{d\left(\delta\mathbf{S}_i\cdot \overline{\mathbf{S}}_k\right)}{dt}
    = -\delta\mathbf{S}_i\cdot \overline{\mathbf{S}}_j - \delta\mathbf{S}_j\cdot \overline{\mathbf{S}}_i\\
    \pm \Big[\overline{\mathbf{S}}_k\cdot\overline{\mathbf{S}}_l\ \left(\delta\mathbf{S}_i\cdot \overline{\mathbf{S}}_m + \delta\mathbf{S}_m\cdot \overline{\mathbf{S}}_i\right)\\
    -\overline{\mathbf{S}}_k\cdot\overline{\mathbf{S}}_m \left(\delta\mathbf{S}_i\cdot \overline{\mathbf{S}}_l + \delta\mathbf{S}_l\cdot \overline{\mathbf{S}}_i\right)\Big].
  \end{multline}
\end{subequations}
Note that the terms in the last two lines of each of these equations (i.e. the terms in the square
brackets, which result from the interactions of spin $\mathbf{S}_i$ with the spins in the left triangle
in Fig.~\ref{fig:cluster_labels}) appear rather cumbersome. This is, however, deceptive since half of them always vanish. For each of the triaxial ground states either $\mathbf{S}_j=\pm \mathbf{S}_l$ while $\mathbf{S}_k=\pm \mathbf{S}_m$ or $\mathbf{S}_j=\pm \mathbf{S}_m$ while $\mathbf{S}_k=\pm \mathbf{S}_l$. Consequently, either $\mathbf{S}_j\cdot \mathbf{S}_l=\pm 1$ while $\mathbf{S}_j\cdot \mathbf{S}_m=0$ or  $\mathbf{S}_j\cdot \mathbf{S}_l= 0$ while $\mathbf{S}_j\cdot \mathbf{S}_m=\pm 1$, with the opposite holding for the dot products involving $\mathbf{S}_k)$. Hence one of the two terms in the square brackets in each equation~ always vanishes while the second one becomes similar to the first line with the signs depending on the specific triaxial state.

Solving the system of these linearized coupled EOMs results in the the magnon spectrum, which is then integrated over the Brillouin zone to obtain the quantum correction to the classical ground state energy.

\subsection{Invariance of semiclassical dynamics under $\pi$-weathervane rotations}
\label{sec:weathervane_EOM}

Having established the semiclassical equations of motion \eqref{eq:EOM_scalar}, we can readily investigate the effects of a weathervane rotation, transforming one triaxial state into another, on the spin-wave spectrum (without having explicitly calculated the latter just yet). While we will specifically focus on the rotations of the affected spins by $\pi$, which we dub $\pi$-weathervane modes, we note that any rotations by multiples of $\pi/2$ preserve the triaxial nature of the state.

Given the unavoidably cumbersome general form of the linearized EOMs in Eqs.~\eqref{eq:EOM_scalar}, we can greatly simplify the analysis by focusing solely on the explicit form of the right-hand side terms in the first line of each of these equations (while avoiding the need to explicitly spell out the other terms). Thus, for conceptual clarity, let us begin by investigating ``truncated'' EOMs given by the first line in each of the Eqs.~\eqref{eq:EOM_scalar} alone; physically they correspond to the dynamics of spin $\mathbf{S}_i$ due to its interaction with spins $\mathbf{S}_j$ and $\mathbf{S}_k$ only:
\begin{subequations}\label{eq:EOM_toy}
  \begin{eqnarray}\label{eq:EOM_toy_a}
    \frac{d\left(\delta\mathbf{S}_i\cdot \overline{\mathbf{S}}_j\right)}{dt}
    &=& \delta\mathbf{S}_i\cdot \overline{\mathbf{S}}_k + \delta\mathbf{S}_k\cdot \overline{\mathbf{S}}_i,
    \\
    \label{eq:EOM_toy_b}
    \frac{d\left(\delta\mathbf{S}_i\cdot \overline{\mathbf{S}}_k\right)}{dt}
    &=& -\delta\mathbf{S}_i\cdot \overline{\mathbf{S}}_j - \delta\mathbf{S}_j\cdot \overline{\mathbf{S}}_i.
  \end{eqnarray}
\end{subequations}

A weathervane rotation involving a particular lattice triangle will invariably rotate two of its spins around the axis defined by the direction of the third spin in the ground state. Therefore we need to consider two distinct cases, with the spin $\mathbf{S}_i$ either being or not being affected by the weathervane rotation.

Let us first consider the latter case, i.e. the case where the new ground state is obtained from the original one by rotating both $\overline{\mathbf{S}}_j$ and $\overline{\mathbf{S}}_k$ around the direction of $\overline{\mathbf{S}}_i$ by $\pi$ resulting in $\overline{\mathbf{S}}_j\to -\overline{\mathbf{S}}_j$, $\overline{\mathbf{S}}_k\to -\overline{\mathbf{S}}_k$ (while $\overline{\mathbf{S}}_i\to \overline{\mathbf{S}}_i$). Consequently, the EOMs~(\ref{eq:EOM_toy}) now read
\begin{subequations}\label{eq:EOM_scalar_after_WV1}
  \begin{eqnarray}\label{eq:EOM_scalar_after_WV1a}
    - \frac{d\left(\delta\mathbf{S}_i\cdot \overline{\mathbf{S}}_j\right)}{dt}
    &=& - \delta\mathbf{S}_i\cdot \overline{\mathbf{S}}_k + \delta\mathbf{S}_k\cdot \overline{\mathbf{S}}_i,
    \\
    \label{eq:EOM_scalar_after_WV1b}
    -\frac{d\left(\delta\mathbf{S}_i\cdot \overline{\mathbf{S}}_k\right)}{dt}
    &=& \delta\mathbf{S}_i\cdot \overline{\mathbf{S}}_j - \delta\mathbf{S}_j\cdot \overline{\mathbf{S}}_i.
  \end{eqnarray}
\end{subequations}
Note that $\delta \mathbf{S}_n$ are dynamical variables for which we may recycle the same notations as those used in Eq.~(\ref{eq:EOM_toy}). The question before us is whether Eqs.~(\ref{eq:EOM_scalar_after_WV1})
can in fact be reduced back to Eqs.~(\ref{eq:EOM_toy}) by a change of variables. This can indeed be achieved by redefining the components of $\delta \mathbf{S}_n$ for each spin participating the $\pi$-weathervane rotation in such a way that its component along the axis of the weathervane rotation changes sign while its component normal to that direction does not. Applied to Eqs.~(\ref{eq:EOM_scalar_after_WV1}) this entails $\delta\mathbf{S}_j\cdot \overline{\mathbf{S}}_i\to - \delta\mathbf{S}_j\cdot \overline{\mathbf{S}}_i$ and $\delta\mathbf{S}_k\cdot \overline{\mathbf{S}}_i\to - \delta\mathbf{S}_k\cdot \overline{\mathbf{S}}_i$ while all dot products containing  $\delta\mathbf{S}_i$ remain unaffected by this change of variables. After doing so, Eqs.~(\ref{eq:EOM_scalar_after_WV1}) become identical to Eqs.~(\ref{eq:EOM_toy}) up to an overall sign of all terms.

The other setting to consider is the case where the $\pi$-weathervane mode involves site $i$ and one of the two remaining sites, $j$ or $k$. Without any loss of generality, let us assume that it is site $j$ and hence the weathervane rotation is performed around the direction of $\overline{\mathbf{S}}_k$ resulting in $\overline{\mathbf{S}}_i\to -\overline{\mathbf{S}}_i$, $\overline{\mathbf{S}}_j\to -\overline{\mathbf{S}}_j$. Consequently, the EOMs~(\ref{eq:EOM_toy}) now read
\begin{subequations}\label{eq:EOM_scalar_after_WV2}
  \begin{eqnarray}\label{eq:EOM_scalar_after_WV2a}
    -\frac{d\left(\delta\mathbf{S}_i\cdot \overline{\mathbf{S}}_j\right)}{dt}
    &=&  \delta\mathbf{S}_i\cdot \overline{\mathbf{S}}_k - \delta\mathbf{S}_k\cdot \overline{\mathbf{S}}_i,
    \\
    \label{eq:EOM_scalar_after_WV2b}
    \frac{d\left(\delta\mathbf{S}'_i\cdot \overline{\mathbf{S}}_k\right)}{dt}
    &=& \delta\mathbf{S}_i\cdot \overline{\mathbf{S}}_j + \delta\mathbf{S}_j\cdot \overline{\mathbf{S}}_i.
  \end{eqnarray}
\end{subequations}
The aforementioned change of variables now implies ${\delta\mathbf{S}_i\cdot \overline{\mathbf{S}}_k\to - \delta\mathbf{S}_i\cdot \overline{\mathbf{S}}_k}$ while preserving $\delta\mathbf{S}_j\cdot \overline{\mathbf{S}}_i$ and $\delta\mathbf{S}_i\cdot \overline{\mathbf{S}}_j$ (along with $\delta\mathbf{S}_k\cdot \overline{\mathbf{S}}_i$ since $\delta\mathbf{S}_k$ is not affected by the change of variables; the weathervane mode does not involve site $k$).
Once again, after such a  change of variables, Eqs.~(\ref{eq:EOM_scalar_after_WV2}) become identical to Eqs.~(\ref{eq:EOM_toy}) up to an overall sign of all terms.

Extending this proof from the ``truncated'' EOMs~(\ref{eq:EOM_toy}) to the original ones given by Eqs.~(\ref{eq:EOM_scalar}) is trivial, if notationally cumbersome since we also have to keep track of which of the additional sites, $l$ and $m$, is affected by the weathervane rotation (and in the first case considered here both of them may or may not be a part of the weathervane mode). The upshot remains: for each site $n$ participating in the $\pi$-weathervane rotation, a change of variables that redefines the components of $\delta \mathbf{S}'_n$  in such a way that its component along the axis of that rotation changes sign while its component normal to that direction does not,  makes the EOMs for the ground state obtained by said weathervane rotation equivalent to those for the original state.

Several comments are now in order. Firstly, the effects of flipping the ground state direction of a particular spin $\overline{\mathbf{S}}_n \to - \overline{\mathbf{S}}_n$ followed by the aforementioned change of variables for $\delta \mathbf{S}_n$ is mathematically equivalent to rotating the spin axes for ${\mathbf{S}}_n = \overline{\mathbf{S}}_n + \delta \mathbf{S}_n$ by $\pi$ around the axis defined by its immediate neighbors \emph{participating} in the $\pi$-weathervane mode (i.e. \emph{not} around the axis of the weathervane rotation itself!). That is, if the axis of the actual $\pi$-weathervane rotation transforming one GS into another is the $z$-axis, the linearized EOMs for the new state are equivalent to those for the old state after rotating the spin axes for the participating $x$-aligned spins by $\pi$ around the $y$-axis, and for the  $y$-aligned spins around the $x$-axis.   In other words, the effect of physically changing the ground state combined with the additional change of variables is entirely encompassed by the canonical transformation of the spin variables in the original ground state with a natural consequence: an invariance of the spin-wave spectrum.

Secondly, it is instructive to check why such an invariance of the spin-wave spectrum holds only for the proper $\pi$-weathervane rotations and not for any flipping of a ground state configuration of spins belonging to a loop containing no sharp corners. This is because such a ground state transformation would invariably involve a spin with two participating neighbors aligned along two different axes in their ground state configuration. Consequently, one can no longer define a canonical transformation for this spin -- their mutual neighbor -- that establishes the equivalence of the respective EOMs.

\subsection{Spin-wave spectra}
\label{sec:uniform_spectra}

We now turn to the central step in the discussion of the quantum order-by-disorder mechanism: the actual calculation of the spin-wave spectrum for the different ground states of our model \eqref{eq:hamiltonian}. We will again focus on the the \qz and \rthree states, but remind the reader that neither the \qz nor the \rthree state is a true ground state of the Hamiltonian~(\ref{eq:hamiltonian}) in the {\em entire} range of $-1\leq \lambda\leq 1$.

\subsubsection*{Spin-wave spectra of the \qz state}
\begin{figure}[t]
\centering
    \includegraphics[width=\linewidth]{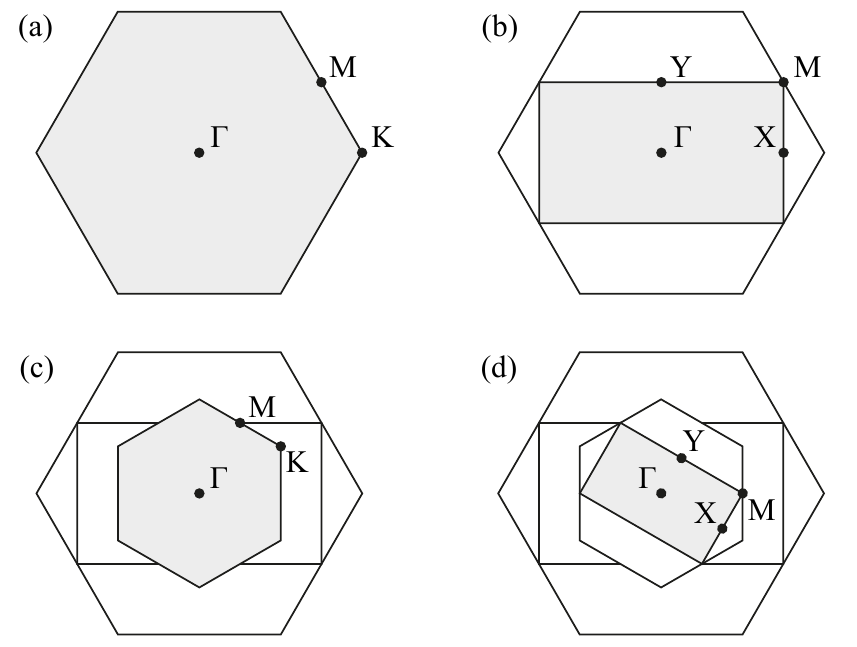}
    \caption{{\bf Brillouin zones and high-symmetry points} for the kagom\'e lattice and its enlarged magnetic unit cells:
    			(a) 3-spin unit cell; (b) 6-spin unit cell; (c) 9-spin unit cell; (d) 18-spin unit cell.}
	\label{fig:BZs}
\end{figure}

The \qz state, specifically, is a ground state of our model with uniform chirality, $0 \leq \lambda\leq 1$, as well as at $\lambda = -1$ (but not for $-1< \lambda < 0$). Furthermore, the pure chiral cases $\lambda = \pm 1$ are special: the \qz ground states are not unique due to the existence of an additional $\mathbb{Z}_2$ degree of freedom; see discussion in Sec.~\ref{sec:chiral_gs}. In our discussion here we focus on the ground states with the smallest possible unit cell, i.e. the shortest periodicity.  The chiral Hamiltonian with uniform chirality, $\lambda = 1$, allows for the assignment of the $\mathbb{Z}_2$ variables that does not enlarge the unit cell of the three-color \qz state; this state is then adiabatically connected to the unique (up to a global permutation of colors) three-color \qz ground state of the mixed model ($0<\lambda<1$), all the way to the KHAFM  ($\lambda=0$). By contrast, the shortest-period assignment of the $\mathbb{Z}_2$ variables in the pure chiral case with staggered chirality, $\lambda = - 1$,
doubles the size of the unit cell of the three-color \qz state. Consequently, the Brillouin zone in this case is a half of that in the cases of $0\leq\lambda\leq 1$; the Brillouin zones are shown in Figs.~\ref{fig:BZs}a and~\ref{fig:BZs}b, respectively.

\begin{figure*}
\centering
    \includegraphics[width=\linewidth]{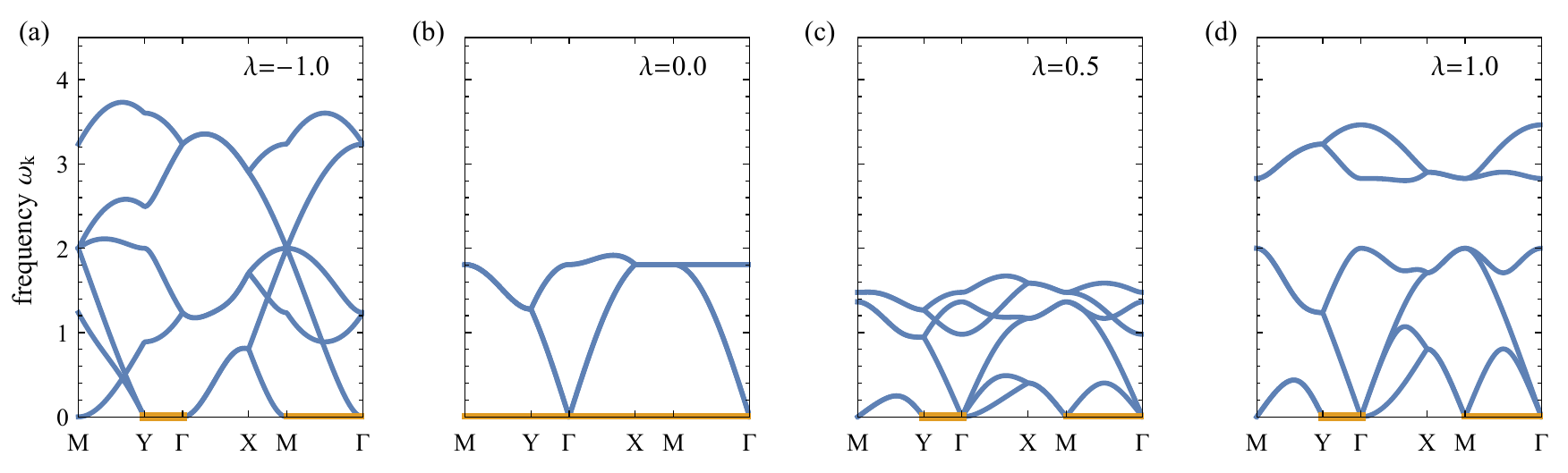}
    \caption{{\bf Spin-wave spectra of the \qz state} for slices through the Brillouin zone in Fig.~\ref{fig:BZs}b at different values of $\lambda$: (a) the staggered chiral model ($\lambda=-1$); (b) the KHAFM ($\lambda=0$); (c) the mixed uniform chiral model ($\lambda=0.5$); (d) the uniform chiral model ($\lambda=1$). Zero-energy bands are shown in orange.
    The spectra are drawn along a path through the high-symmetry points of the reduced Brillouin zone of Fig.~\ref{fig:BZs}b.
    Note that in order to represent the full set of bands of these \qz states, we have defined them on a {\em doubled} 6-spin unit cell for better comparison later with the \qz state when $\lambda = -1$ (which necessarily has a 6-site unit cell).}
	\label{fig:double_qz}
\end{figure*}

\begin{figure*}
	\centering
	\includegraphics[width=\linewidth]{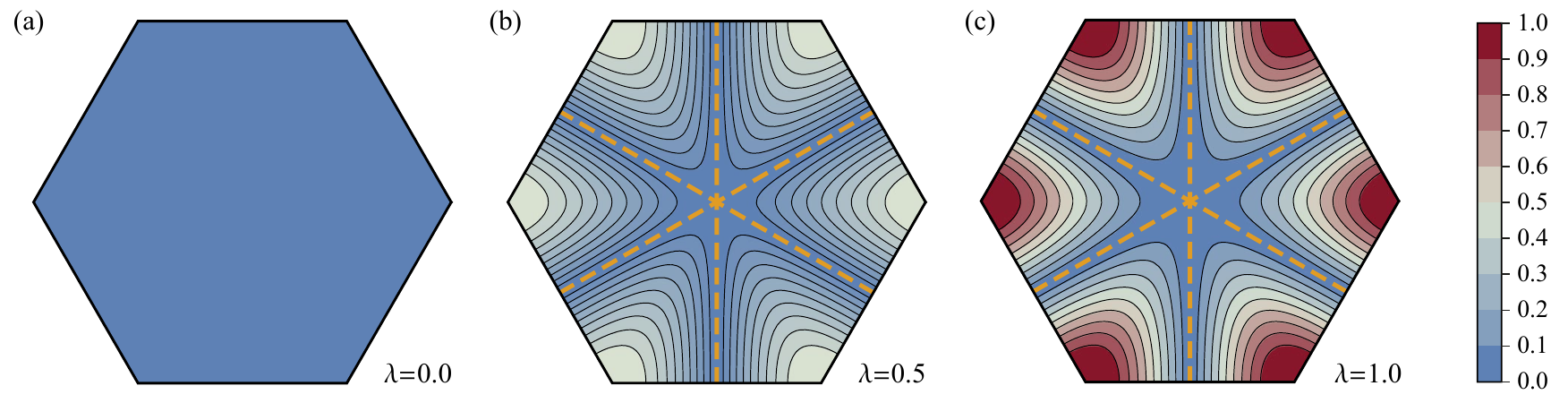}
    \caption{The {\bf lowest band of the \qz state} as contour plots for different values of $\lambda$: (a) the KHAFM ($\lambda=0$); (b) equal mixing ($\lambda=0.5$); (c) the uniform chiral model ($\lambda=1$). While the lowest band is extensively degenerate at $\lambda=0$, its degeneracy is partly lifted to subextensive lines of momenta (indicated by dashed orange lines) at $\lambda>0$. The contour plots span the Brillouin zone shown in Fig.~\ref{fig:BZs}a which corresponds to a 3-spin unit cell.}
	\label{fig:qz_contours}
\end{figure*}

\begin{figure*}
\centering
    \includegraphics[width=\linewidth]{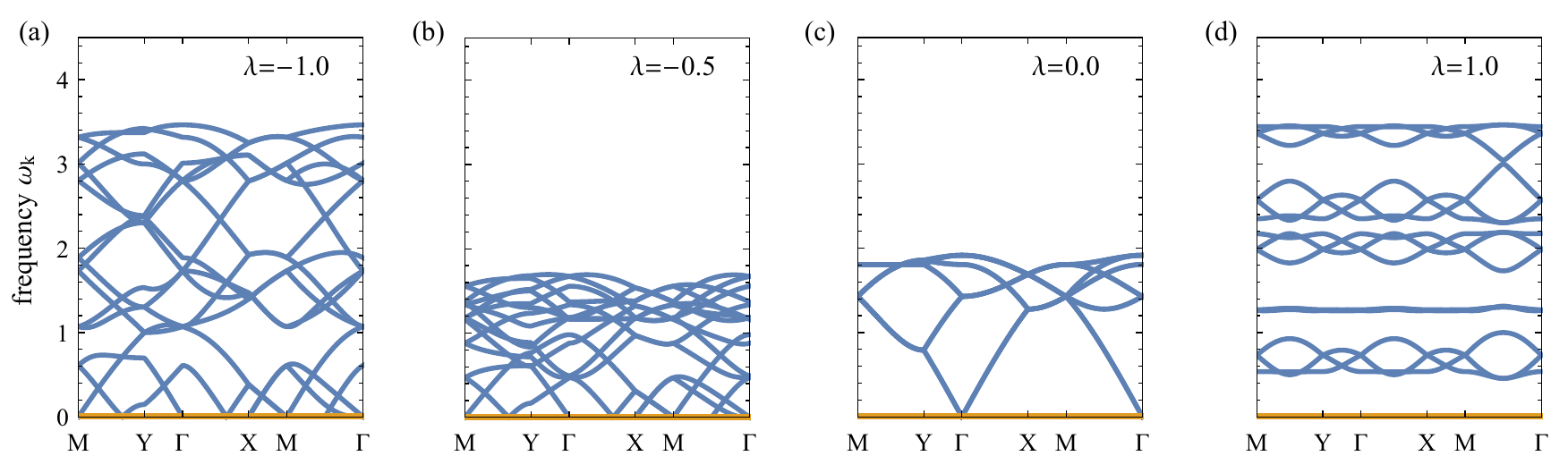}
    \caption{{\bf Spin-wave spectrum of the \rthree state} for slices through the Brillouin zone in Fig.~\ref{fig:BZs}d at different values of $\lambda$: (a) the staggered chiral model ($\lambda=-1$); (b) the mixed staggered chiral model ($\lambda=-0.5$); (c) the KHAFM ($\lambda=0$); (d) the uniform chiral model ($\lambda=1$). Zero-energy bands are shown in orange.}
	\label{fig:slices_root3}
\end{figure*}

Fig.~\ref{fig:double_qz} presents slices through the spin-wave spectra of the \qz state made along straight lines connecting high-symmetry points marked in Figs.~\ref{fig:BZs}a and ~\ref{fig:BZs}b. An important feature of these spectra is the existence of zero-energy \emph{lines} marked in orange, which can be visualized as ``grooves'' along high-symmetry lines in the contour plots of the lowest energy band shown in Fig.~\ref{fig:qz_contours}. For $\lambda\neq 0$ such a behavior is fully expected based on the analysis of harmonically soft modes in Sections~\ref{sec:soft_modes_q0} and \ref{sec:soft_modes_mixed}.
The Heisenberg point, $\lambda=0$, is special in that the lowest band flattens, becoming a zero-energy band -- see Fig.~\ref{fig:qz_contours}a. This is, once again,  consistent with the properties of classical soft modes in the KHAFM; see Section~\ref{sec:ToBD_KHAFM}.

\subsubsection*{Spin-wave spectra of the \rthree state}
We now turn to the \rthree ground states. From the discussion in Sec.~\ref{sec:groundstates}, we know that the \rthree three-color state is a ground state of a model with staggered chirality, $-1 \leq \lambda\leq 0$, as well as at $\lambda = 1$, but not in the interval $0< \lambda < 1$. Analogously to the \qz case above, the pure chiral cases $\lambda = \pm 1$ are special due to the existence of an additional $\mathbb{Z}_2$ degree of freedom. However, this time it is the chiral Hamiltonian with staggered chirality, $\lambda = - 1$ that allows for the assignment of the $\mathbb{Z}_2$ variables that does not enlarge the unit cell of the three-color \rthree state whereas the unit cell must be doubled in the case of uniform chirality, $\lambda = 1$. Since the unit cell of the three-color \rthree state already consists of nine sites, it increases to 18 for $\lambda = 1$. The corresponding Brillouin zones are shown in Figs.~\ref{fig:BZs}c and~\ref{fig:BZs}d.

High-symmetry line cuts through the spin-wave spectra are shown in Fig.~\ref{fig:slices_root3}. The remarkable feature of these spectra is that in contrast to the  \qz ground states, the lowest energy spin-wave excitations in the \rthree ground states {\em always} form a flat zero-energy band. This is, once again, fully consistent with the analysis of soft modes in Sections~\ref{sec:soft_modes_root3} and \ref{sec:soft_modes_mixed}. As a side remark, we note a curious feature of the spectrum in Fig.~\ref{fig:slices_root3}d obtained for the pure chiral model with uniform chirality, $\lambda = 1$. The zero energy flat band in this case does not touch any other bands, which might seem surprising in light of the counting arguments presented in Ref.~\cite{Bergman2008}. However, that argument is based on the existence of zero modes that wind around the torus. Accounting for those modes exceeds the capacity of one band and hence necessitates band touching. As has been argued in Sec.~\ref{sec:soft_modes_root3}, the number of independent harmonically soft modes in a \rthree ground state of a purely chiral model is equal to that of the hexagons of the kagom\'{e} lattice, i.e. the number of kagom\'{e} unit cells, which account for one filled band. In the absence of ``topological'' zero modes, the arguments of Ref.~\cite{Bergman2008} are not applicable; similar behavior has been recently observed in another frustrated magnetic system~\cite{Bilitewski2018}.

subsection{Zero-point energies}
\label{sec:integration}
The key difference between the the thermal and quantum order-by-disorder mechanisms is that the former is driven solely by the harmonically soft modes whereas the latter is a consequence of \emph{all} harmonic excitations.  All spin-wave bands contribute quantum zero-point energy corrections to the energy of the ground state. In integral form
the harmonic energy correction per spin can be expressed as
\begin{equation}
\Delta\mathcal{E}_\text{harm}= \frac{1}{2}\frac{1}{n_b A_\text{BZ}}\sum_b \int_\text{BZ} d^2\mathbf{k}\; \omega_b(\mathbf{k})=\frac{1}{2}\langle\omega\rangle \,,
\end{equation}
where $n_b$ is the number of bands (i.e. the number of spins in a magnetic unit cell) and $A_\text{BZ}$ is the area of the Brillouin zone.

Numerical estimates  \footnote{
The values presented were obtained by a quadratic fit to the sequence of trapezoidal rule approximations.
All trapezoidal rule results from values of $h>h_\text{min}$ were used for the fit, which was very good, having a variance smaller than the machine precision.
} of this harmonic energy correction
obtained for the \qz and \rthree states of the uniform and staggered chiral models with the smallest unit cells for each type of the three-color states and each chirality are given in Table~\ref{table:eh}.
Evidently, the harmonic correction to the classical energy of the \qz states exceeds that of the \rthree states in both models.
This suggests a quantum order-by-disorder mechanism stabilizing the \rthree state for both uniform and staggered chiral Hamiltonians. But there remains a barrier to asserting this selection with confidence: the impossibility of testing all classical ground states corresponding to the same three-color state but different arrangements of $\mathbb{Z}_2$ variables (see Sec.~\ref{sec:chiral_gs}).
For the  \rthree states this is not an issue, since all possible $\mathbb{Z}_2$ arrangements can be obtained from one another by a sequence of $\pi$ weathervane modes and hence result in exactly the same harmonic zero-point energy (as discussed above in Sec.~\ref{sec:weathervane_EOM}).
But the same does not hold for the \qz three-color states. This motivates us to construct an effective low-energy theory for these semiclassical harmonic corrections.

\begin{table}[t]
\centering
\begin{tabular}{l l l c l c l }
\hline
state & $\quad  \lambda \quad $ & $\Delta\mathcal{E}_\text{harm}$ \\
\hline \hline
\qz & \quad $+1$ \quad & 0.82260(4)  &     \\
\rthree & \quad $+1$\quad  &  0.82182(5) &      \\
\hline
\qz & \quad $-1$ \quad& 0.83111(1)  &    \\
\rthree & \quad $-1$ \quad & 0.80558(5)  &
\end{tabular}
\caption{\textbf{Semiclassical correction} to classical ground state energy per spin of various ground states for $\lambda = \pm1$}
\label{table:eh}
\end{table}

\subsection{An effective low-energy theory for QObD}
\label{sec:effective}

As we have seen in the previous sections, the presence of the $\mathbb{Z}_2$ degree of freedom that describes the direction of a spin along a particular axis in a triaxial ground state introduces an interesting
new dimension that differentiates pure chiral models from the pure Heisenberg case. Specifically, we have shown that a certain class of $\mathbb{Z}_2$ transformations -- the $\pi$ weathervane rotations -- do not affect the spin wave spectrum and hence do not lift the degeneracy of the corresponding ground states via quantum order-by-disorder mechanism. On the other hand, a generic $\mathbb{Z}_2$ transformation (i.e. flipping all spins along a loop containing no sharp corners) does not posses this property. Since one cannot possibly test all possible $\mathbb{Z}_2$ arrangements numerically, one can attempt to develop an effective low-energy theory that captures the effect of those transformations, e.g. in the spirit of Refs.~[\onlinecite{Hizi2005,Henley2006,Hizi2006}] where such a theory was developed for pyrochlore magnets.

\subsubsection{An effective Hamiltonian}
\label{sec:effectivehamiltonian}

The effective low-energy theory discriminating between different triaxial ground states of the chiral model should in general be a $\mathbb{Z}_3\times\mathbb{Z}_2$ theory, with the $\mathbb{Z}_3$ sector corresponding to the assignment of one of the three axes to each site and the $\mathbb{Z}_2$ sector corresponding to the alignment or antialignment of a particular spin along its local axis. The largest terms in an effective Hamiltonian should therefore be those enforcing the correct chirality for each triangle. This can be implemented by an antiferromagnetic Potts interaction between neighboring $\mathbb{Z}_3$ degrees of freedom and an additional three-body term that favors the arrangement of $\mathbb{Z}_2$ degrees of freedom that results in the correct sign of chirality for each triangle, given its arrangement of the $\mathbb{Z}_3$ degrees of freedom. The second term couples the $\mathbb{Z}_3$ and $\mathbb{Z}_2$ sectors, making the general theory inescapably complicated. We can, however, try simplifying our task by ``freezing'' the $\mathbb{Z}_3$ sector and considering a residual $\mathbb{Z}_2$ theory in each sector. This could be a meaningful exercise if the energy difference between different $\mathbb{Z}_3$ sectors -- i.e. the energy associated with $\pi/2$ weathervane rotations connecting these sectors -- exceeds that between the states related by $\mathbb{Z}_2$ transformations alone. In what follows, we describe our attempt to construct such an effective $\mathbb{Z}_2$ theory for two specific $\mathbb{Z}_3$ sectors corresponding to the \qz and \rthree states.
\begin{figure}[t]
\centering
    \includegraphics[width=0.9\columnwidth]{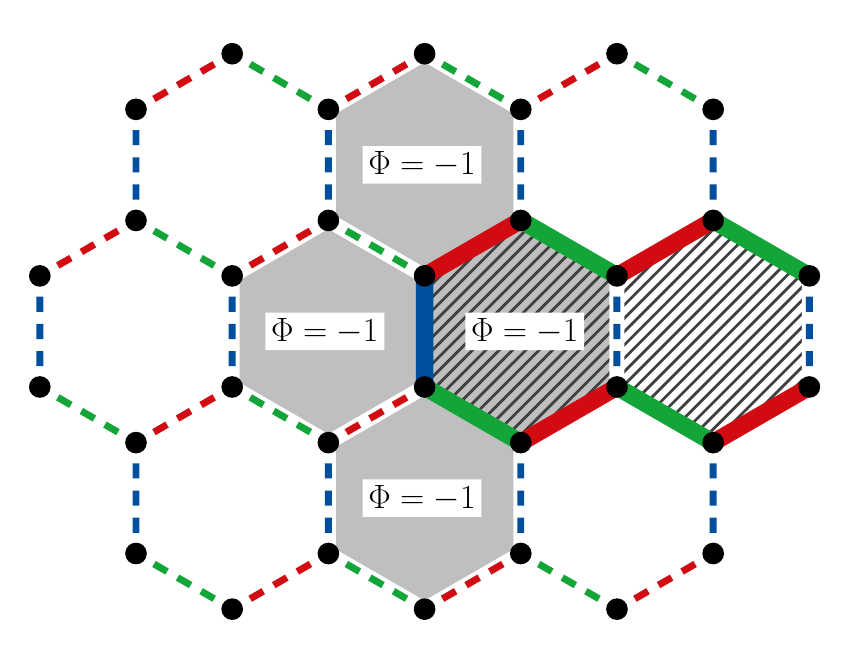}
    \caption{{\bf A $\mathbb{Z}_3\times\mathbb{Z}_2$ configuration} corresponding to one of the $q=0$ arrangements of $\mathbb{Z}_3$ bond variables which are indicated by the three colours. The $\mathbb{Z}_2$ variables correspond to a bond being dashed or solid. For a uniform chiral Hamiltonian ($\lambda=1$) with the assignment of $x$, $y$ and $z$ to red, green and blue respectively, $\sigma^z=1$ is indicated by a dashed line whereas a solid line corresponds to $\sigma^z=-1$. In a ground state configuration solid lines must form closed loops; a segment of such a loop is shown here. All non-winding loops can be generated from an ``empty'' state by a sequence of plaquette flips that reverse all $\mathbb{Z}_2$ variables around a plaquette; in this example the flipped plaquettes are hatched. A plaquette flip is not a gauge-like transformation in the $q=0$ state and should result in energy change (see text for details). An end of a line of flipped plaquettes can be detected using the $\mathbb{Z}_2$  flux operator defined in Eq.~(\ref{eq:flux}); four plaquettes with negative flux (i.e. odd parity of solid bonds) are shaded grey in this example.}
	\label{fig:gauge_q0}
\end{figure}

It is natural for the effective $\mathbb{Z}_2$ theory to be a gauge theory since the degeneracy is not lifted by $\mathbb{Z}_2$ transformations corresponding to the $\pi$ weathervane rotations. In an effective gauge theory such transformations -- gaugelike transformations in the parlance of Refs.~[\onlinecite{Hizi2005,Henley2006,Hizi2006}]  --  become genuine gauge transformations. In the traditions of $\mathbb{Z}_2$ gauge theories, we formulate it on the honeycomb lattice where sites of the honeycomb correspond to the centers of kagom\'{e} triangles and, consequently, the original sites of the kagom\'{e} lattice now correspond to the bonds of the honeycomb lattice. (It is customary to associate gauge fields with bonds rather than sites of a lattice.) The chirality constraint then translates into a vertex (AKA star) term in the effective $\mathbb{Z}_2$ Hamiltonian: given a fixed $\mathbb{Z}_3$ arrangement, the vertex term in the $\mathbb{Z}_2$ sector will favor either an even or odd number of $\mathbb{Z}_2$ ``spins'' (denoted henceforth as $\sigma$) to be $+1$ around a particular vertex. Therefore
\begin{equation}\label{eq:vertex}
\mathcal{H}_v = \sum_v J_v\prod_{i\in v}\sigma_i^z \,,
\end{equation}
where the sum is taken over all vertices $v$ of the honeycomb lattice while the product is taken over all bonds $i$ adjacent to a particular vertex. The coupling constant $J_v=\pm J$ depends on the $\mathbb{Z}_3$ arrangement around that vertex. For instance, for a staggered chiral Hamiltonian ($\lambda=-1$), the \rthree state shown in Fig.~\ref{fig:KHAFM_gs} would correspond to either all $J_v=J>0$ or all $J_v=-J$ depending on whether the assignment of axes $x$, $y$ and $z$ to the three colours $A$, $B$ and $C$ forms a right- or a left-handed triad. The uniform chiral Hamiltonian ($\lambda=1$), on the other hand, would result in a staggered assignment of $J_v=\pm J$ for the two sublattices of the honeycomb lattice. The same uniform Hamiltonian would, however, result in a uniform
assignment of $J_v=\pm J$ for the \qz $\mathbb{Z}_3$ sector; an example of both $\mathbb{Z}_3$ and $\mathbb{Z}_2$ variable assignments that satisfy the chirality constraint in this case is shown in Fig.~\ref{fig:gauge_q0}.

Beginning with the \rthree state, we observe that every honeycomb plaquette is ``flippable'': negating all $\mathbb{Z}_2$ degrees of freedom corresponds to a $\pi$ weathervane rotation of the original spins around a corresponding hexagon of the kagom\'{e} lattice. Therefore, the product $\prod_{i\in \hexagon}\sigma_i^x$ can be thought of as a pure gauge transformation in the effective $\mathbb{Z}_2$ theory for the \rthree states. It commutes with the vertex terms, and since all arrangements of the $\mathbb{Z}_2$ degrees of freedom that satisfy the vertex terms result in the same semiclassical zero-point energy, the effective $\mathbb{Z}_2$ theory for the \rthree states contains no other terms. (We shall revisit this point later.)

\begin{figure}[b]
\centering
    \includegraphics[width=0.9\columnwidth]{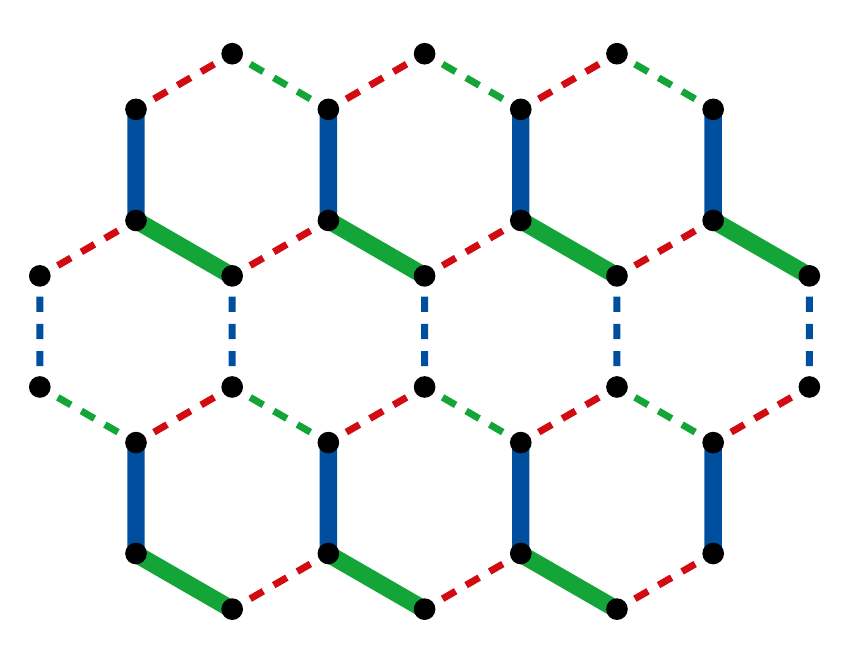}
    \caption{{\bf A $\mathbb{Z}_3\times\mathbb{Z}_2$ ground-state configuration for a staggered chiral Hamiltonian.} It corresponds to one of the $q=0$ arrangements of $\mathbb{Z}_3$ bond variables which are indicated by the three colours. The $\mathbb{Z}_2$ variables correspond to a bond being dashed or solid. Again, $x$, $y$ and $z$ axes are associated with red, green and blue respectively. Pseudospin $\sigma^z=1$ is indicated by a dashed line whereas a solid line corresponds to $\sigma^z=-1$. All plaquettes in this configuration have flux $\Phi=-1$.}
	\label{fig:gauge_q0_staggered}
\end{figure}

\begin{figure}
    \centering
    \includegraphics[width=\linewidth]{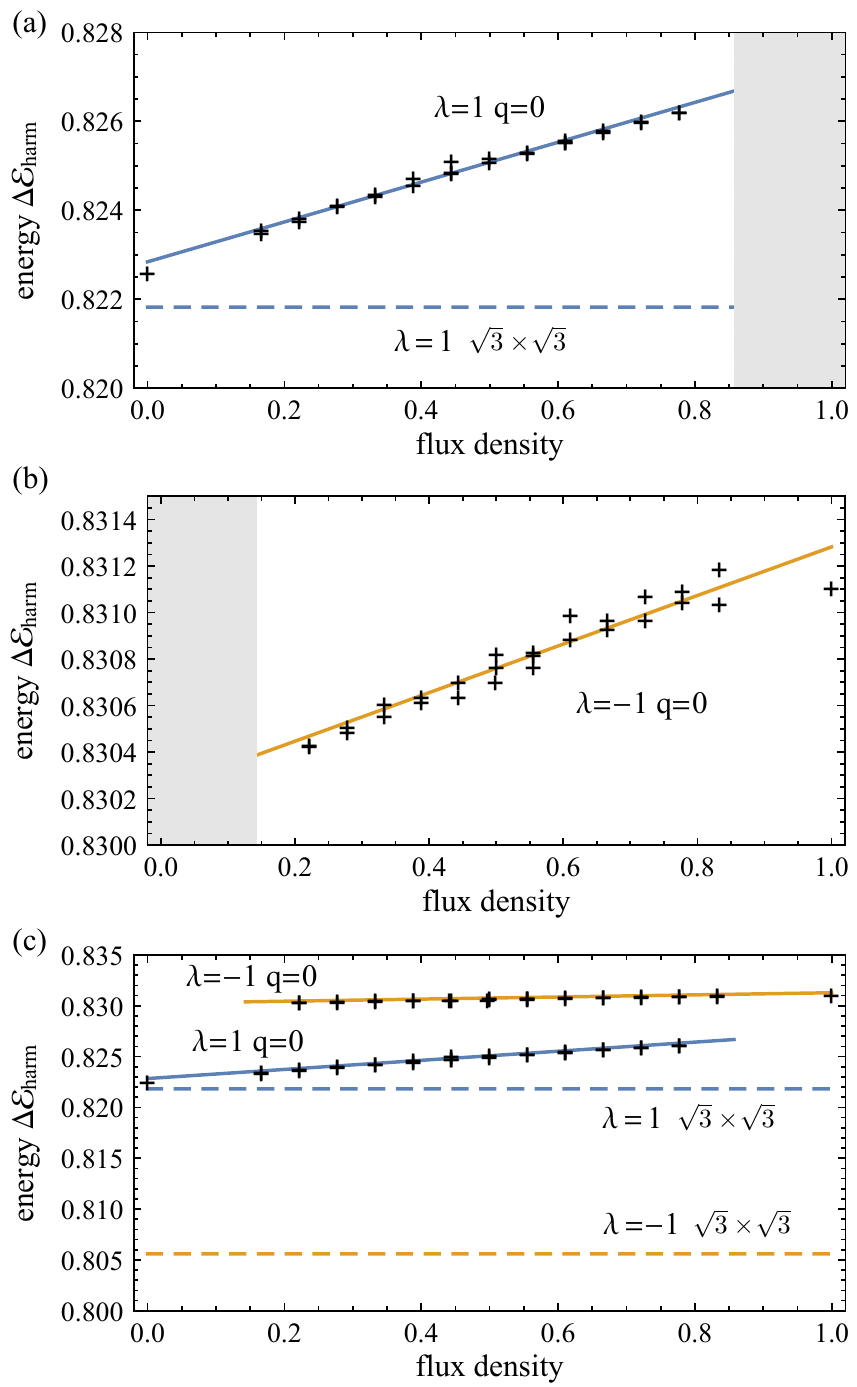}
    \caption{\textbf{Semiclassical harmonic correction} of the ground state energy per spin for different \qz states of (a)~the uniform chiral model and (b)~the staggered chiral model. (c)~Energy of the different \qz states shown in subpanels (a) and~(b) in comparison to the energy of the \rthree state.}
\label{fig:qzflips}
\end{figure}

That is clearly not the case for the \qz states since the plaquette flip operator $\prod_{i\in \hexagon}\sigma_i^x$ no longer corresponds to a $\pi$ weathervane rotation of the original spins. However, it still commutes with the vertex terms and consequently the effective Hamiltonian must contain other terms, with which it does not commute. One such operator is the plaquette flux operator which we define as
\begin{equation}\label{eq:flux}
  \Phi_{\hexagon} = \prod_{i\in \hexagon}\sigma_i^z \,.
\end{equation}
While the flux operator commutes with the plaquette flip operator acting on the same plaquette, they \emph{anticommute} when acting on two neighboring plaquettes.

The heuristic reason to expect that the effective Hamiltonian will contain flux terms is as follows. We know that the only possible weathervane modes in the \qz state are those corresponding to a rotation of the (original) spins situated on a straight line of the kagom\'{e}. (With periodic boundary conditions, this line wraps around the torus.) Two $\pi$ weathervane rotations of spins situated on two neighboring parallel lines are translated into flipping the entire line of adjacent hexagons in the effective theory. Clearly, such an operation cannot change the energy of the state. Incidentally, this operation also leaves all fluxes intact since any hexagon that neighbors a flipped plaquette in fact neighbors exactly two of those and hence its flux is flipped twice. On the other hand, if we were to terminate a line of flipped plaquettes, the flux would change in four hexagons adjacent to last plaquette of the line as can bee seen in Fig.~\ref{fig:gauge_q0}. Bending a line of flipped hexagons also entails changing the flux in two plaquettes immediately adjacent to the bend. Therefore, the flux defined by Eq.~(\ref{eq:flux}) is not only sensitive to the deviations of plaquette flip sequences from those corresponding to the $\pi$ weathervane rotations, it detects those deviations \emph{locally}. Hence, including flux operators into the effective Hamiltonian will preserve the locality of the theory. This intuition also finds support in the construction of the effective low-energy theory for the pyrochlore magnets~\cite{Hizi2005,Henley2006,Hizi2006}.

Focusing on the $\mathbb{Z}_2$ flux defined by Eq.~(\ref{eq:flux}) has additional heuristic appeal. We have already seen that the arrangement of $\mathbb{Z}_2$ degrees of freedom around each vertex depends on whether the $\mathbb{Z}_3$ degrees of freedom have been arranged to form a right- or left-handed triad. For instance, in the example shown in Fig.~\ref{fig:gauge_q0}, the association of red, green and blue with $x$, $y$ and $z$ axes led to the vertex term in Eq.~(\ref{eq:vertex}) favoring an odd number of negative pseudospins $\sigma_z$ around each vertex. Had we associated these colours with  $x$, $z$ and $y$ axes instead, the vertex term would have to favor an even number of negative pseudospins -- or odd number of positive pseudospins. Graphically, this would result in either replacing all dashed lines in Fig.~\ref{fig:gauge_q0} with solid ones, and vice versa, or by changing the designation of a solid line to denote $\sigma_z=1$ (instead of $\sigma_z=-1$). Crucially, this would not change the $\mathbb{Z}_2$ flux through each plaquette! In other words, the notion of $\mathbb{Z}_2$ flux is invariant with respect to one's choice between energetically equivalent $\mathbb{Z}_3$ sectors.

One interesting aspect to observe is that while the $q=0$ state of the uniform chiral Hamiltonian with the smallest magnetic unit cell (a unit cell for its $\mathbb{Z}_2$ pseudospins) is flux-free, the opposite holds for such a state in the case of the staggered chiral Hamiltonian. For the staggered Hamiltonian, the smallest magnetic unit cell is twice the size of a lattice unit cell and such a state has $\Phi_{\hexagon} = -1$ through each plaquette, as can be seen in Fig.~\ref{fig:gauge_q0_staggered}. Any pseudospin flips around a closed, non-winding loop will reduce the net negative flux.

While it is natural to expect that the energy will depend both on the total flux and its spatial distribution -- we generally expect that plaquettes with negative flux interact with one another -- here we pursue a less ambitious programme and simply investigate numerically the harmonic corrections to the classical GS energy for several arrangements of flux insertion into the most uniform (as far as $\mathbb{Z}_2$ degrees of freedom are concerned) ground states. This allows us to estimate the couplings in the conjectured effective Hamiltonian
\begin{equation}
\label{eq:PHAFMeff}
\mathcal{H}_\text{eff}^\text{q=0}=E_0 + \frac{\mu}{2} \sum_{\hexagon}\left(1- \Phi_{\hexagon}\right) + \mathcal{H}_\text{int},
\end{equation}
where $\mu$ serves as the ``chemical potential'' for (negative) flux whereas $\mathcal{H}_\text{int}$ encompasses interaction between fluxes -- the term we have not attempted tackling analytically but whose magnitude we can estimate by considering different spatial distributions of fluxes.

Figure~\ref{fig:qzflips} shows the harmonic correction to the classical ground state energy evaluated for $q=0$ states with different densities of $\mathbb{Z}_2$ flux in a system of $6\times 6$ hexagons. Each flux sector is represented by two different configurations with the difference between their energies providing an estimate for the magnitude of the interaction term. The average slope allows us to estimate $\mu$ and extrapolate the lines through the entire range of flux densities. Note that the maximum flux density for the case of uniform chirality ($\lambda=1$) is $6/7$, i.e. 6 out of every 7 plaquettes can have $\Phi_{\hexagon}=-1$, wheres for the staggered case ($\lambda=-1$), the minimum density is $1/7$. Note that in both uniform and staggered cases, (negative) $\mathbb{Z}_2$ flux is energetically costly albeit the cost differs dramatically in the uniform case and staggered cases; the similarity of slopes in Figures~\ref{fig:qzflips}(a) and \ref{fig:qzflips}(b) is misleading and is an artefact of different rescaling of energy along the vertical axes. Figure~\ref{fig:qzflips}(c) shows these plots presented on the same scale with the dashed lines representing the energy of the \rthree states (which is independent of flux). Therefore it appears very convincing that the lowest energy $q=0$ states are indeed the states with the lowest flux density. Furthermore, in cases of both uniform and staggered chiral interactions the energies of these states are higher than those of the \rthree
states and consequently the latter are energetically selected. Note that the gap between the flux-free $q=0$ state and the \rthree state of the uniform chiral Hamiltonian is minuscule -- see also Table~\ref{table:eh}. The gap is significantly larger in the staggered case, which makes us confident that the $q=0$ state with the lowest flux density (1/7) is not a GS contender even though we did not evaluate its energy explicitly; all numerical results presented in this section have been obtained on a lattice consisting of $6 \times 6 = 36$ hexagons (or, equivalently, 108 original spins) with periodic boundary conditions, which does not accommodate a state with the flux density of 1/7.

We conclude that both the QObD and TObD mechanisms result in the selection of the same \rthree states. Note that while common, such a scenario is not guaranteed; we are aware of at least one example where different orders are stabilized by thermal and quantum fluctuations~\cite{Toth2010}.

\subsubsection{The \rthree states: A semiclassical spin liquid}
\label{spin_liquid}

Our analysis strongly suggest that semiclassical corrections favor \rthree states over \qz classical ground states for both uniform and staggered chiral Hamiltonian, i.e. the same states are stabilized by both thermal and quantum order by disorder. We have further argued that different \rthree states obtained by reversing all $\mathbb{Z}_2$ degrees of freedom around a plaquette -- a plaquette flip -- have the same energy (at least in the harmonic order). Therefore the effective Hamiltonian for the $\mathbb{Z}_2$ pseudospins can contain only vertex terms -- see Eq.~(\ref{eq:vertex}); the plaquette flips are equivalent to gauge transformations in this language. However, thinking of them as pure gauge transformations is misleading since the pseudospins correspond to physical, measurable degrees of freedom. A plaquette flip corresponds to a physical $\pi$ weathervane rotation of all original spins around a given kagom\'e hexagon. It is natural to expect that such a process will emerge in a $1/S$ expansion as a higher order term~\cite{von_Delft1992}. With the inclusion of such a term in the low-energy description, the effective Hamiltonian becomes
\begin{equation}
\label{eq:Kitaev}
\mathcal{H}_\text{eff} = \sum_v J_v\prod_{i\in v}\sigma_i^z + J_p \sum_p \prod_{j\in p}\sigma_j^x \,,
\end{equation}
where the second sum is taken over all plaquettes of the honeycomb lattice.
This, of course, is the well-known Hamiltonian of Kitaev's toric code~\cite{Kitaev2003} -- an archetypal model system whose ground states manifest the $\mathbb{Z}_2$ topological order. Note that the signs of individual vertex coupling constants $J_v$ are immaterial for the spectral properties of this Hamiltonian and the topological nature of its ground states.

The emergence of this effective Hamiltonian present a fascinating possibility: the formation of a semiclassical $\mathbb{Z}_2$ spin liquid stabilized by a quantum order-by-disorder mechanism. Furthermore, we have already seen that thermal fluctuations also select the \rthree state (while being oblivious to the $\mathbb{Z}_2$ sector). Consequently, in a finite-size system the states stabilized by the TObD mechanism are the ground states of the same effective Hamiltonian~(\ref{eq:Kitaev}). (The finite-size requirement is a consequence of the Mermin--Wagner theorem; any triaxial order entails breaking of the global O(3) symmetry which is impossible in the thermodynamic limit at any finite temperature.)   Weathervane modes will mix those states producing a Gibbs state that at a small but finite temperature will be described by the density matrix corresponding to Kitaev's toric code on a finite system at the temperature low enough to prevent ``electric'' excitations but high enough to allow proliferation of ``magnetic'' excitations~\cite{Castelnovo2007}. (The specific temperature regime corresponds to  $(J_p/\ln N)<T<(J_v/\ln N)$ where $J_v$ and $J_p\ll J_v$ are the coupling constants associated with vertex and plaquette operators in the toric code.) In this regime, the von Neumann entropy still has a topological contribution of  $\ln 2$, i.e. a half of the topological entropy of the ``fully-fledged'' toric code~\cite{Castelnovo2007}.


\section{Disucssion}
\label{sec:conclusions}

To put our analysis of the chiral kagom\'e model into context, let us compare its physics to the celebrated KHAFM, one of the archetypal examples of a frustrated magnetic system that has extensively many  classical ground states. A special subset of those states, the coplanar states, play an important role in its low-temperature physics: these states become selected by a TObD mechanism at low but non-zero temperature.
The chiral kagom\'e model considered in this manuscript possesses a similarly large ground state degeneracy. A subclass of these ground states -- the triaxial states -- closely resemble the coplanar states of the KHAFM. In fact, they can be continuously connected to the coplanar states via the mixed parameter regime where both Heisenberg and chiral couplings are present (as illustrated in Fig.~\ref{fig:mixed_gs_energy}).
Despite the similarities, there are also some important differences that lead to interesting physical consequences:
Firstly, the resolution of these vast, but accidental ground state degeneracies  varies distinctly between the two models. This is because the number of harmonically soft spin wave modes -- the key ingredient in the order-by-disorder mechanism -- varies between different triaxial states in the chiral model whereas it is the same for the coplanar GSs of the KHAFM. Consequently, the selection of states by either thermal (ToBD) or quantum fluctuations (QoBD) is markedly different: whenever the \rthree state -- the state with the largest number of soft modes --  is a ground state of the chiral model, it is unambiguously selected.
While it appears that the KHAFM also selects \rthree correlations in the limit of low temperature~\cite{Chern2013}, the underlying mechanism is weaker, as it only takes place beyond the harmonic description. As a result, the \rthree order parameter in the KHAFM does not saturate for $T\rightarrow 0$.

Note that, in addition, the fluctuations also stabilize the order-by-disordered regime even for the mixed model where the ground state itself is ordered in the \qz state.
There, the higher energy of the  \rthree state is  compensated by the entropic contribution to the free energy of its soft modes.
We refer to this phenomenon as ``proximate order by disorder'', drawing a loose analogy to the notion of a proximate spin liquid \cite{Banerjee2016,Banerjee2017,Revelli2020} in systems where thermal fluctuations render spin liquid phenomenology visible above an ordered ground state.

In passing we note that the formation and entropic stabilization of triaxial ground states for the classical model in the presence of chiral interactions might also be an essential ingredient in making a connection to the ground states of the quantum model. For it has been argued
that strong quantum fluctuations (going beyond their more gentle role considered in the QoBD mechanism) will lead to a melting
of such non-coplanar order into chiral spin liquids \cite{Hickey2017} as observed, at least for the case of uniform chiralities, in the
quantum model \cite{Bauer2014}.

But perhaps the most interesting distinct feature of the classical chiral model with regard to the KHAFM is the additional $\mathbb{Z}_2$ degree of freedom associated with triaxial states. The order-by-disorder selection of \rthree triaxial states does not discriminate between compatible arrangements of the $\mathbb{Z}_2$ degrees of freedom, which may be thought of as emergent gauge degrees of freedom. The constraints imposed on these degrees of freedom turn out to be the same as those in the Kitaev toric code, i.e. a paradigmatic model of intrinsic $\mathbb{Z}_2$ topological order. Adding appropriate dynamics compatible with these constraints could thus result in the formation of genuine topological order -- a rare example of topological order stabilized by fluctuations.\\


\begin{acknowledgments}
The authors are grateful to J.~T.~Chalker, F.~Mila, A.~Paramekanti,
J.~G.~Rau and S.~L.~Sondhi for comments and discussions.
The Cologne group acknowledges partial funding from the Deutsche Forschungsgemeinschaft (DFG, German Research Foundation)
-- project grants 277101999 and 277146847 -- through SFB 1238 (project C02) and
within the CRC network TR 183 (project A04).
This work was in part supported by the Deutsche Forschungsgemeinschaft under grants SFB 1143 (project-id 247310070) and the cluster of excellence ct.qmat (EXC 2147, project-id 390858490).
KS is grateful for the hospitality of the MPIPKS, Dresden and KITP, Santa Barbara, the latter supported in part by the National Science Foundation under Grant No.~NSF PHY-1748958.
The numerical simulations were performed on the CHEOPS cluster at the RRZK Cologne
and on the JUWELS cluster at the Forschungszentrum Juelich.
Our numerical simulations have made use of the ALPS/ALEA library~\cite{Albuquerque2007,Bauer2011,Alpswebsite}.
\end{acknowledgments}


%

\end{document}